\documentclass[twocolumn,twocolappendix]{aastex701}

\usepackage{aas_macros}
\usepackage{amsfonts}
\usepackage{float}
\usepackage{amsmath,amssymb}
\usepackage{natbib}    
\usepackage{hyperref} 
\usepackage{graphicx} 
\usepackage{color}
\usepackage{verbatim}
\usepackage{enumitem}
\usepackage{natbib}

\usepackage[linesnumbered,ruled]{algorithm2e}

\newcommand{\proptosim}{\mathrel{\vcenter{
 \offinterlineskip\halign{\hfil$##$\cr
 \propto\cr\noalign{\kern2pt}\sim\cr\noalign{\kern-2pt}}}}}
\renewcommand{\b}[1]{\boldsymbol{#1}}
\newcommand{\unit}[1]{{\rm #1}}

\newcommand{\cm}{\unit{cm}}
\newcommand{\m}{\unit{m}}
\newcommand{\g}{\unit{g}}
\newcommand{\G}{\unit{G}}
\newcommand{\K}{\unit{K}}
\newcommand{\km}{\unit{km}}

\newcommand{\eV}{\unit{eV}}

\newcommand{\s}{\mathrm{s}}

\newcommand{\ang}{\ensuremath{\mathrm{\AA}}}

\newcommand{\B}{\mathbf{B}}     
\newcommand{\E}{\mathbf{E}}     
\newcommand{\J}{\mathbf{J}}     
\renewcommand{\v}{\mathbf{v}}   
\newcommand{\kb}{k_\mathrm{B}}

\renewcommand{\d}{\mathrm{d}}

\newcommand{\e}{\mathrm{e}}

\newcommand{\p}{\mathrm{p}}     

\renewcommand{\O}{\mathrm{O}}     
\newcommand{\A}{\mathrm{A}}     
\renewcommand{\H}{\mathrm{H}}     
\renewcommand{\P}{\mathrm{P}}     
\newcommand*\chem[1]{\ensuremath{\mathrm{#1}}}

\usepackage{bm}
\usepackage{amsmath}
\usepackage{amssymb}
\usepackage{graphicx} 
\usepackage{float} 
\usepackage{subfigure}
\usepackage{enumerate}
\usepackage{booktabs}
\usepackage{listings}
\usepackage{makecell}
\usepackage{enumitem} 
\usepackage{placeins}
\usepackage{afterpage}

\begin{document} 

\title{Interstellar Dust-Catalyzed Hydrogen Formation
  Enabled by Nuclear Quantum Effects}

\author[orcid=0009-0009-7727-2403,gname='Xiaolong',
sname='Yang']{Xiaolong Yang} 
\affiliation{School of Materials Science and Engineering,
  Peking University, Beijing 100871, China} 
\email{1900017821@pku.edu.cn}

\author[gname=Lile, sname='Wang']{Lile Wang} 
\affiliation{Kavli Institute for Astronomy and Astrophysics,
  Peking University, Beijing 100871, China} 
\affiliation{Department of Astronomy, School of Physics,
  Peking University, Beijing 100871, China} 
\email[show]{lilew@pku.edu.cn}

\author[gname=Di,sname=Li]{Di Li}
\affiliation{Department of Astronomy, Tsinghua University,
  Beijing 100084, China} \email{dili@tsinghua.edu.cn}

\author[sname=Xu,gname=Shenzhen]{Shenzhen Xu}
\affiliation{School of Materials Science and Engineering,
  Peking University, Beijing 100871, China}
\email{xushenzhen@pku.edu.cn}
\correspondingauthor{Lile Wang}

\begin{abstract}
  Molecular hydrogen (\chem{H_2}) is one of the key
  chemical species that controls and shapes a wide spectrum
  of astrophysical processes from galaxy evolution to planet
  formation. Although catalyzation on dust grain surfaces is
  the dominant formation channel of \chem{H_2} in the
  interstellar medium, its efficiency across $20-200~\K$ has
  remained not fully understood. Here, using multiscale
  simulations combining \textit{ab-initio}-level machine
  learning force fields, constrained path-integral Monte
  Carlo, and kinetic Monte Carlo, we perform a systematic,
  quantum-mechanical study of the full \chem{H_2} formation
  sequence, including hydrogen adsorption, diffusion,
  association and desorption. We explicitly consider the
  decoupling of gas and dust temperatures, making our
  results applicable to photon-dominated regions (PDRs) and
  dense cold clouds. Our results show that on the bare,
  crystalline surfaces studied here (graphitic and silicate
  grains), physisorbed hydrogen is negligible, and nuclear
  quantum effects (NQEs) in chemisorbed hydrogen atoms are
  essential for efficient formation at low temperatures,
  overcoming the classical Boltzmann suppression. This work
  presents a quantitative NQEs-inclusive study on silicate
  surfaces (exemplified by enstatite) and graphitic grains,
  revealing surface-specific adsorption behavior. These
  findings provide a first-principles quantum foundation for
  interstellar \chem{H_2} formation, complementing empirical
  multipliers, and enable new observational constraints on
  dust composition and molecular cloud evolution. The
  framework also extends to other astrochemical reactions on
  dust grains under full NQEs.
\end{abstract}


\section{Introduction}
\label{sec:intro}

Molecular hydrogen (\chem{H_2}) is the most abundant
diatomic molecule in the interstellar medium (ISM) and
serves as an important component in a wide range of related
astrophysical processes, including galaxy formation,
molecular cloud evolution, star formation, and even planet
formation \citep[e.g.][]{shull1982araa, gnedin2009apj,
  christensen2012mnras}. It acts as the primary coolant in
gravitationally collapsing clouds and participates in key
astrochemical networks that drive interstellar chemistry
\citep{Herbst2001review}. While several gas-phase formation
channels exist, such as the three-body reaction
($3\chem{H}\rightarrow \chem{H_2} + \chem{H}$) or the
\chem{H^-} pathway
($\chem{H^-} + \chem{H} \rightarrow \chem{H_2} + e^-$),
these require physical parameters (densities or ionization
fractions) that deviate significantly from conditions in
typical molecular clouds (see e.g.\ \citealt{Draine_book}).
Instead, \citet{Hollenbach1971ApJ} proposed and established
the physical picture of \chem{H_2} formation on grain
surfaces.  

Although grain-catalyzed formation has become the {\it de
  facto} standard theory of \chem{H_2} formation in
molecular clouds, significant gaps remain in our
quantitative understanding of the process
\citep{Vidali2013ChemRev, Wakelam2017MolAs}. Most
importantly, \chem{H_2} has been observed to form
efficiently across a wide temperature range, from
$\sim 20~\K$ \citep{Cazaux2002ApJ, GoldsmithLi2005ApJHINSA,
  Vidali2013ChemRev} through $\gtrsim 200~\K$
\citep{GriecoEtAl2023NA}, within and outside the Galaxy
\citep[e.g.][]{GnedinKravtsovApJ2010,
  FeldmannEtAl2023MNRAS}.  The physical mechanisms that
support such efficiency, nevertheless, remain debated.
Given the non-negligible energy barriers
($\Delta E_{\rm a}\gtrsim 0.5~\eV$) for hydrogen atom
diffusion and association, the low-temperature efficiency
would classically be suppressed by more than 50 orders of
magnitude by the Boltzmann factor
[$\exp(-\Delta E_{\rm a}/\kb T)$] at $T\sim 20~\K$. Although
amendments have been proposed to resolve this issue, for
example, by assuming reactions between physisorbed and
chemisorbed hydrogen atoms, such models still cannot explain
efficient formation at relatively high temperatures
($T \gtrsim 150~\K$), where rapid desorption of physisorbed
atoms (usually with $\Delta E_{\rm a} \sim 10^{-2}~\eV$ and
desorption timescales $\lesssim 10^{-12}~\s$) also strongly
inhibits \chem{H_2} formation.  To address these issues
physically, it is essential to study the formation channel
starting from chemisorbed hydrogen atoms, which requires the
inclusion of nuclear quantum effects (NQEs) including
zero-point energy and quantum tunneling, to allow hydrogen
atoms to overcome energy barriers at low temperatures while
not suffering from rapid desorption at high temperatures.
Nevertheless, a quantitative and comprehensive model that
analyzes the entire sequence of \chem{H_2} formation is
still needed, particularly with the increasing desire to
interpret observed molecular fractions in terms of dust
grain properties and astrophysical environments using
advanced observational studies
\citep[e.g.][]{GriecoEtAl2023NA, Goldsmith2025ApJ}.
Moreover, interstellar dust grains are primarily composed of
silicates (such as olivine and enstatite) and carbonaceous
materials (including graphite, amorphous carbon, and
polycyclic aromatic hydrocarbons; see also
\citealt{Draine2003ARAA}).  Simulating the NQEs-influenced
formation process on both carbonaceous and silicate
materials could therefore provide further astrophysical
insights into this fundamental process.

There are two regimes of \chem{H_2} formation that
  should be noted: on bare grain surfaces, where
  chemisorption is relevant, and on ice-covered grains,
  where physisorption dominates. The latter is a
  well-established field: \chem{H_2} forms efficiently on
  amorphous solid water and other ice analogs at low
  temperatures \citep{1999Takahashi, 2008Watanabe,
    2013Watanabe, Wakelam2017MolAs}, and subsurface ice
  chemistry drives molecular evolution in dark clouds
  \citep{2010Kalv} (see reviews by
  \citealt{Vidali2013ChemRev, 2013Watanabe,
    Wakelam2017MolAs}). Our present work focuses on bare
  grains, where classical Boltzmann suppression requires
  NQEs to restore efficiency, instead of challenging the
  physisorption-dominated pathway on ice-covered grains,
  which could remain as one of the primary channels in
  dense, cold environments.

 The upgrades in computing methods and their
  physical implications is also relevant in the
  explorations. Earlier NQE studies relied on
  one-dimensional tunneling with simplified barrier shapes
  or transfer matrices \citep{1987Ando, 1994Walker},
  neglecting multi-dimensional effects like corner cutting
  \citep{1977Marcus}, in both ice \citep{2008Watanabe,
    2012Oba, 2013Minissale, 2014Congiu, 2017Senevirathne}
  and bare-grain contexts \citep{2014Rimola, 2015Rimola,
    2016Rimola, 2019Kerkeni, 2022Kerkeni,
    Tong2024JPCC}. More recently, semiclassical instanton
  theory has provided valuable tunneling estimates for
  astrochemistry \citep{2010Goumans, 2016Lamberts,
    2016Meisner, 2017Meisner, 2017Senevirathne, 2019Meisner,
    2021Enrique-Romero, HanEtAl2022JPCL}. In these
  approaches, crossover temperature $T_{\rm c}$
  ($\sim100$--$300~\K$ for H reactions on forsterite and
  pyroxene; \citealt{2015Rimola, 2016Rimola, 2019Kerkeni})
  marks the onset of deep tunneling. However, the
  path-integral framework seamlessly spans both regimes:
  NQEs are built into the free energy profiles and rate
  constants without switching approximations. We therefore
  employ full path-integral Monte Carlo (PIMC), averaging
  over all quantum imaginary paths — important when a single
  dominant path is not fully representative — and manage its
  computational cost through machine-learning force fields
  (MLFFs).
  
  Admittedly, alternative mechanisms have been
  proposed for bare grains. \citet{2005A&A...434..599C,
    2005Cuppen, 2006Cuppen} showed via KMC that surface
  roughness and stochastic heating can extend efficient
  \chem{H_2} formation using coordination-dependent
  effective rates. Complementary DFT studies
  \citep{2013Kerkeni, 2015Kerkeni, 2017Kerkeni, 2019Kerkeni,
    2022Kerkeni, 2025Kerkeni, 2014Rimola, 2015Rimola,
    2016Rimola} found H chemisorption barrierless on many
  silicates. However, these remain static or semiclassical
  treatments near $0~\K$; systematic temperature-dependent
  quantum-statistical studies of the full $20$--$200~\K$
  sequence have not been available until now.
  
In this work, KMC (e.g. \citealt{MieEtAl2019FiC})
simulations are applied based on the single-step PIMC data,
enabling research on larger grain surfaces to eliminate
boundary effects. By including all elementary steps in
consistent quantum mechanical computations, our method
features more comprehensive understanding over previous
pioneering works \citep[e.g.][]{HanEtAl2022JPCL}, performing
quantitatively reliable statistical path-integral sampling
over different dust compositions, which is necessary for
accurate NQEs simulations of the entire reaction chain,
including adsorption and desorption.  We expect that the
long-standing astrophysical puzzle, the relatively high
efficiency of dust-catalyzed \chem{H_2} formation across a
wide temperature range ($20~\K \lesssim T \lesssim 200~\K$),
can be consistently resolved with a proper full-procedure
NQEs model.  This letter is structured as follows. The basic
methodologies are briefly described in \S\ref{sec:methods},
and the quantum-mechanical modeling and KMC results for the
formation efficiency are presented in \S\ref{sec:results}
for both carbonaceous and silicate dust grain surface
models. \S\ref{sec:summary} describes the astrophysical
implications of these results. The details of our NQEs
simulations are elaborated in the Appendices.


\section{Methods}
\label{sec:methods}

\begin{figure*}[ht!]
\centering  
\includegraphics[width=0.7\linewidth]{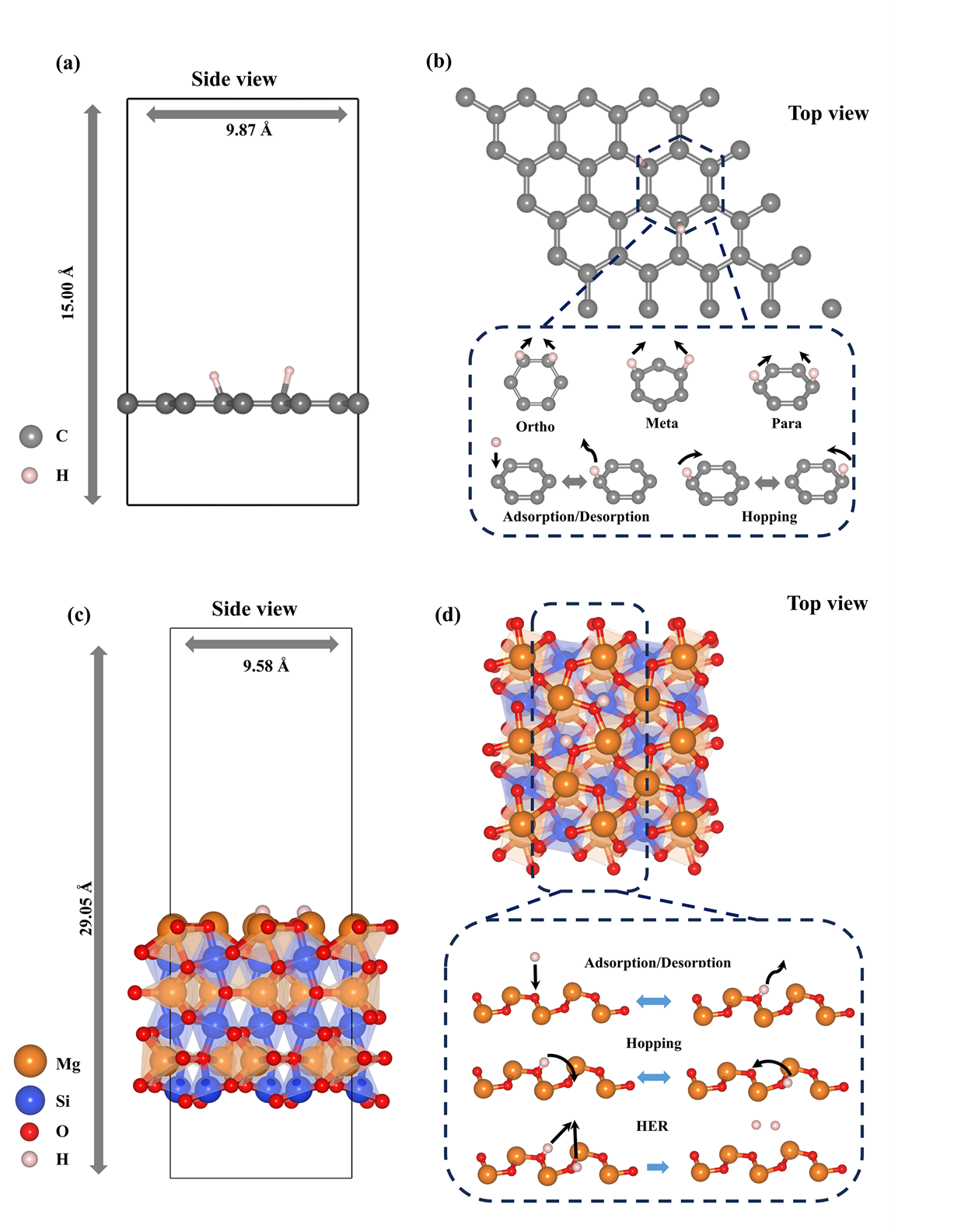}
\caption{(a) Side view of the simulated ($4\times 4$)
  graphene surface with two hydrogen atoms absorbed. (b) Top
  view of the graphene surface and schematic diagrams of
  elementary steps.  (c) Side view of the simulated
  ($2\times 2$) \chem{MgSiO_3} slab model with three layers
  of \chem{Mg_8Si_8O_{24}}
  and the medium layer is fixed
  during calculations. The top surface is terminated with
  Mg-O and the bottom surface is terminated with O-Si-O. (d)
  Top view of the Mg-O termination and schematic diagrams of
  elementary steps investigated in this work.
  \label{fig:structure}}
\end{figure*}

The core of studying \chem{H_2} formation efficiency is to
evaluate the effective activation energy for all elementary
steps in the formation of \chem{H_2} on the surfaces of
carbonaceous and silicate grains. For simplicity and
certainty, we choose graphene for carbonaceous grains
\citep{Chen2017ApJ, HanEtAl2022JPCL, Tong2024JPCC,
  Tong2024CPL}, and enstatites (Pnma-\chem{MgSiO_3})
\citep{MatProj2013} as for silicates \citep{Henning2010araa,
  Molster2002AAa, Molster2002AAb}, and readers are referred to Appendix~\ref{sec:model} for detailed reasoning. We build a
graphene slab with 32 carbon atoms shown in Figure
\ref{fig:structure}(a) with all elementary steps relevant to
\chem{H_2} formation illustrated in Figure
\ref{fig:structure}(b). For the surface of
Pnma-\chem{MgSiO_3}, a three-layer slab model is
constructed with a top surface terminated with Mg-O and a
bottom surface terminated with O-Si-O, shown in Figure
\ref{fig:structure}(c). The primary steps are shown in
Figure \ref{fig:structure}(d). Refer to
Appendix~\ref{sec:model} for details about two slab
models. Hydrogen adsorption is found to be more stable at
the O sites on the Mg-O termination (by $\sim 0.5~\eV$,
refer to Table~\ref{tab:site}), leading to a
quasi-one-dimensional chain structure along which we model
the adsorption and desorption processes, hopping and
two-hydrogen (two-H hereafter) association. Computational
details are shown in Appendix~\ref{sec:dft}.

\begin{figure*}[ht!]
\centering  
\includegraphics[width=0.7\linewidth]{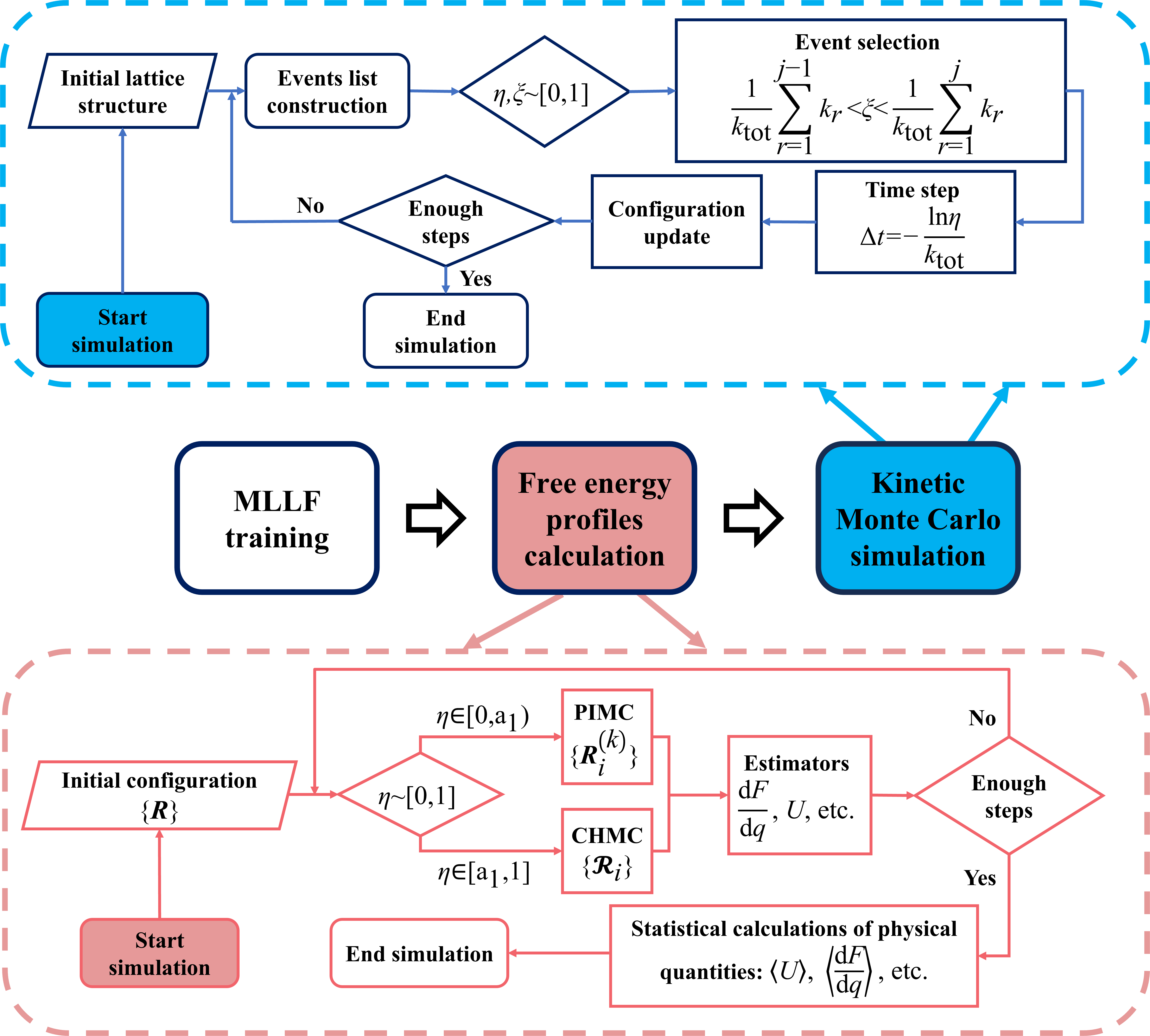}
\caption{Workflow of our multiscale simulation,
  which contains three major modules: MLFF training,
  free energy calculations (highlighted in pink),
  and KMC simulations (highlighted in blue).
  The free energy module uses the CPIHMC method to sample
  configurational space while accounting for NQEs via path
  integrals (Appendix~\ref{sec:free_energy}), and the KMC
  module evolves grain-scale reaction kinetics using rate
  constants derived from the free energy barriers.
  \label{fig:workflow}}
\end{figure*}

The workflow of our multiscale simulation is shown in Figure
\ref{fig:workflow}, which has three modules, MLFFs training
\citep{Lu2021CPCdpmd, WangEtAL2018DPMDKit}, free energy
calculations, and KMC simulations \citep{MieEtAl2019FiC}. In
the MLFFs training module, DP-GEN software
\citep{ZhangEtAl2020DPGEN} is used to train MLFFs based on
data from density functional theory (DFT)
\citep{Kohn1965DFT}. The details of DFT calculations and
MLFFs training are provided in Appendix~\ref{sec:dft} and
\ref{sec:mlffs}.

\afterpage{
\begin{deluxetable}{ccccccc}[h]
\tabletypesize{\scriptsize}
\tablewidth{0pt} 
\tablecaption{Relative energies $E_{\mathrm{site-H}}$ (eV)
  of one hydrogen atom \label{tab:site}} 
\tablehead{
\colhead{Site} &\colhead{} & \multicolumn{2}{c}{Mg-O}&
\colhead{} & \multicolumn{2}{c}{O-Si-O}  
\\
\cline{3-4}\cline{6-7}
\colhead{} &\colhead{} & \colhead{O Site} & \colhead{Mg
  Site} & \colhead{} & \colhead{O Site} & \colhead{Si Site} 
}
\startdata 
 { }& { } & 0 &  1.02  &  { }  & 0.47 & 0.91 \\
\enddata
\tablecomments{Measured at different possible sites of
  \chem{MgSiO_3} surfaces, referred to the energy value of
  \chem{H^*} at the O site of the Mg-O terminated surface.
}
\end{deluxetable}
}

In the free energy calculation module, we employ the
thermodynamic integration (TI) method \citep{Den2000JCP,
  Sprik1998JCP} to calculate free energy profiles
constrained along predefined reaction coordinates (RCs) with
the path integral algorithm \citep{Feynman1948PI} to
consider NQEs of protons by an efficient sampling strategy,
the constrained hybrid Monte Carlo (CHMC) method
\citep{Jin2023JCTC, Sun2025NC}. The definitions of RCs are
elaborated in Appendix~\ref{sec:rc}, while the theories for
free energy calculations are introduced in
Appendix~\ref{sec:free_energy}.

In KMC simulation module, the rate constants of
elementary reaction steps are obtained from free energy
calculations based on the traditional transition state
theory (TST):
\begin{equation}
  k_{\rm TST}=\frac{\kb T}{h} \exp
  (-\beta\Delta F^\ddagger),
\end{equation}
where $k_\mathrm{TST}$ is the reaction rate constant, $h$ is
the Planck constant, and $\Delta F^\ddagger$ is the
activation free energy. For the adsorption reaction, we
modify the prefactor $\kb T/h$ to $nvS$, where
$(n/\cm^{-3})\in \{10^2, 10^4, 10^6\}$ stands for the atomic
hydrogen density, $v = \sqrt{8 \kb T/\pi m}$ is the average
velocity of hydrogen atoms according to the Maxwell
distribution in the space around ISM, and $S$ is the average
area of each adsorption site. Here we admit that TST is an
approximate theory, where the dynamic effect (e.g.,
re-crossing at the dividing surface along a reaction
process) is neglected. We then perform KMC simulations,
which are widely used to study the dynamic properties of a
system and based on the events list built with reaction rate
constants of elementary steps. The readers are referred to
Appendix~\ref{sec:kmc} for details of KMC simulations.


\section{Results}
\label{sec:results}

\subsection{Reduction of effective free energy barriers via NQEs}

We calculate free energy profiles by integrating the mean
force (derivative of free energy with respect to RC) at
different values of RC with $\sim 10^5$-step trajectories.
To investigate the impact of NQEs, the PI algorithm is
implemented in the CHMC method (CPIHMC hereafter)
  with a total number of 64 sampling ``beads'' for paths.
Here we mainly present the results of the graphene case, and
the \chem{MgSiO_3} case is elaborated in
Appendix~\ref{sec:mg_free}.

\begin{figure*}[ht!]
\centering  
\includegraphics[width=0.7\linewidth]
{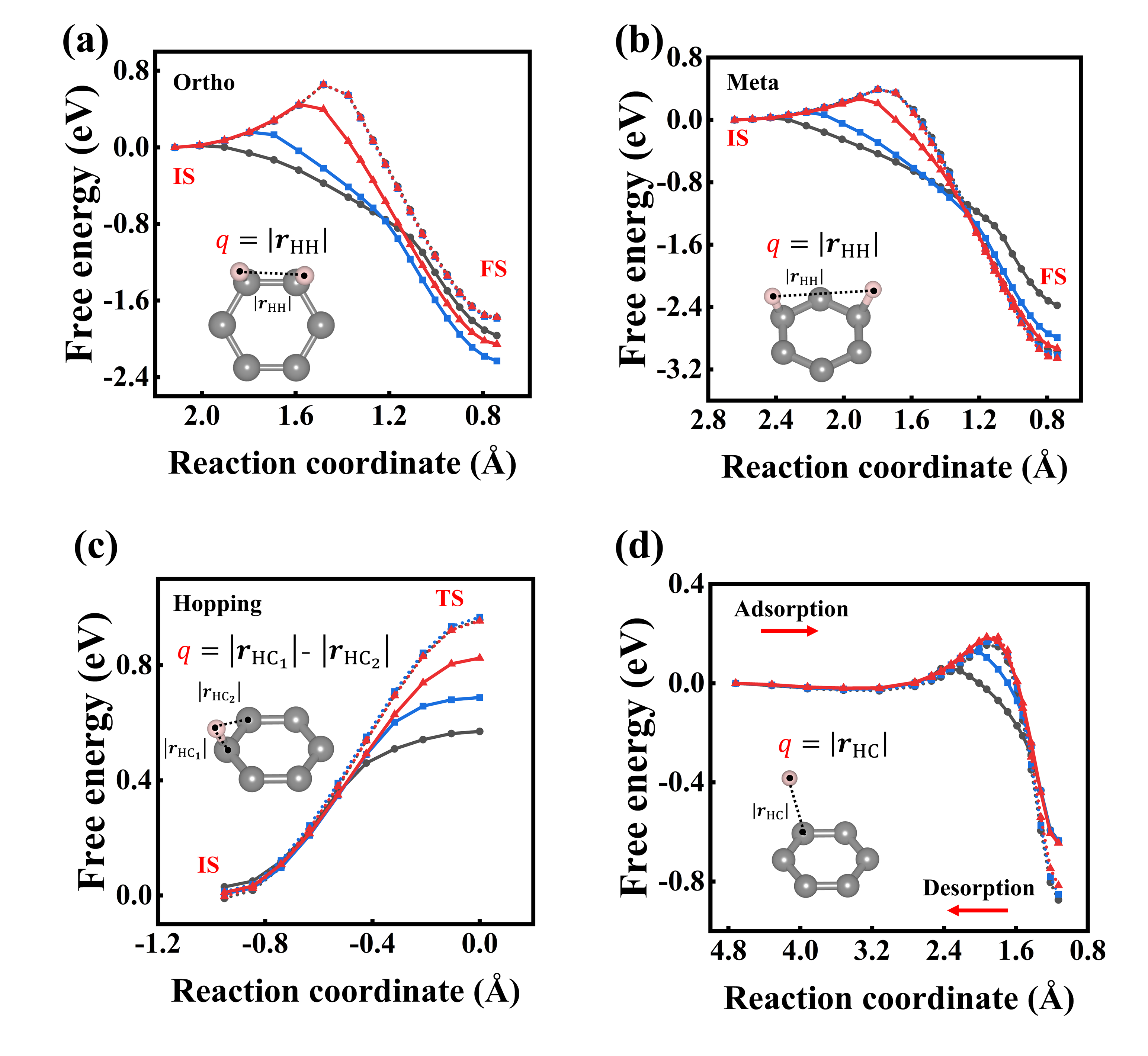}
\caption{Free energy profiles of elementary steps from the
  initial state (IS) to the final state (FS) on the graphene
  surface under quantum (solid lines) and classical (dotted
  lines) situations at temperatures of $50~\K$ (black),
  $100~\K$ (blue), and $200~\K$ (red). The hopping process
  is symmetric from the IS (left) to the FS (right), so we
  just show the part from IS to the transition state
  (TS). Panel (a) for two-H association from an ortho
  configuration (left for IS and right for FS), (b) for wo-H
  association from a meta configuration (left: IS; right:
  FS), (c) for hopping (left: IS; right: TS), and (d) for
  the adsorption and desorption of a hydrogen atom.
  \label{fig:free_energy}}
\end{figure*}

The representative results are presented in Figure
\ref{fig:free_energy} showing the ortho- and meta-channels
of association, while the para-channel is presented in
Appendix~\ref{sec:mg_free} due to its higher barrier and
lower reaction rate at the same temperatures. With
simulations conducted at $50~\K$, $100~\K$, and $200~\K$,
the activation free energies at other temperatures within $50-200~\K$ are
approximately obtained by Lagrange interpolation. Our results reveal
that at low temperatures, considering NQEs the activation barriers for
two-H association on graphene are extremely low, with
barriers dropping below 30~meV at 50~K. While classical
treatment yields negligible hopping and desorption rates,
quantum tunneling significantly accelerates these processes,
even surpassing adsorption under certain conditions. A
similar enhancement is observed on silicate surfaces
(Appendix~\ref{sec:mg_free}), confirming that NQEs are
essential to explain efficient interstellar \chem{H_2}
formation at relevant dust temperatures.
The activation free energies derived from
  the simulations are transferred into reaction rate
  constants through TST. Given that deep tunneling dominates
  the elementary steps at 20~K, we use an extrapolation
  formula to obtain the reaction rate constants at 20~K
  using the data at 50~K and 100~K. The extrapolation
  process is introduced in Appendix~\ref{sec:kmc}.

\subsection{\chem{H_2} formation efficiency calculated by KMC simulations}

We performe KMC simulations to quantify the overall
\chem{H_2} formation rates across astrophysically relevant
temporal and spatial scales with reaction rate constants as
input. There are two kinds of assumptions
  adopted in our simulations, (1) the thermalization
  assumption (suitable for dense, cold environments), and
  (2) the adiabatic assumption (suitable for environments
  such as photodissociation region). Under the adiabatic
  assumption, the gas temperature ($T_\mathrm{gas}$) and the
  dust temperature ($T_\mathrm{dust}$) are decoupled while
  under the thermalization assumption,
  $T_\mathrm{gas} = T_\mathrm{dust}$. The key difference
  between two assumptions lies in the calculation of the
  adsorption rate constant which is discussed in
  Appendix~\ref{sec:assumptions}. The KMC trajectories
(comprising of $\sim 10^5$ steps) are calculated in a range
of temperatures ($20-200~\K$) and
atomic hydrogen densities ($10^2-10^6~\cm^{-3}$), explicitly
incorporating hydrogen adsorption and starting from a
pristine surface. The \chem{H_2} formation rate,
$R_{\chem{H_2},{\rm form}}$, is determined by normalizing
the total number of \chem{H_2} formation events by the
simulated physical time and the surface area of the
surface. From
  $R_{\chem{H_2},{\rm form}}$, the combined $\eta S_{\rm H}$
  is deduced as the normalized value for reaction efficiency
  coefficients \citep{Bai_2009}.  In our simulations, it is approximated that
  the reaction rate constants are all coverage independent,
  which is justified in Appendix~\ref{sec:coverage}. To
  better emulate the amorphous nature of astrophysical dust
  grains, we have also explored the graphitic substrates doped with vacancy defects, and the readers are referred to Appendix~\ref{sec:defect} for details.

\begin{figure*}[ht!]
\centering  
\includegraphics[width=0.9\linewidth]{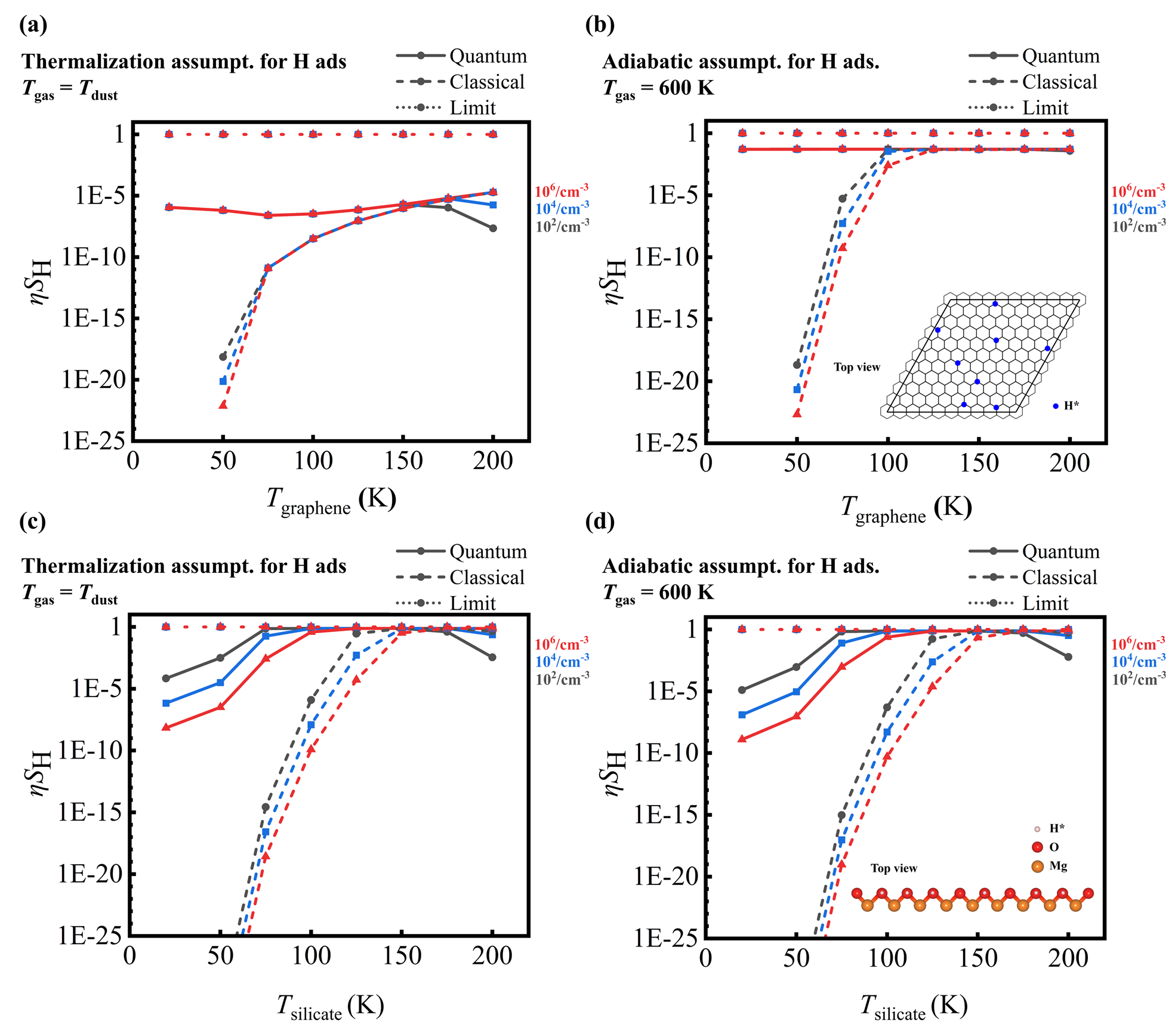}
\caption{Schematic plot of the lattice model employed in the
  KMC simulations and the \chem{H_2} formation efficiency represented by $\eta S_{\rm H}$ under
  different conditions on (a,b) graphene surfaces, and (c,d)
  \chem{MgSiO_3} surfaces. We explicitly show the results of
  quantum (solid lines) and classical (dashed lines) cases
  at different temperatures and surrounding environmental
  hydrogen densities of $10^2~\cm^{-3}$ (black), $10^4~\cm^{-3}$
  (blue), and $10^{6}~\cm^{-3}$ (red). The dotted
  lines represent the efficiency limits. In (a) and (c), we set $T_{\rm gas} = T_{\rm dust}$ (thermalization assumption) while in (b) and (d), we fix $T_{\rm gas} = 600~\K$ (adiabatic assumption).
  \label{fig:kmc}}
\end{figure*}

For the graphene surface, a $10\times 10$ lattice model
composed of $\sim 200$ carbon atoms is constructed to
simulate the catalytic process. Results
under the thermalization and adiabatic assumptions are shown
in Figure \ref{fig:kmc}, panel~a and b respectively. The
results confirm the crucial role of NQEs in the efficient \chem{H_2} formation at low
  temperatures, with much quicker adsorption and association
  than the classical case. Without NQEs, \chem{H_2} formation
is significantly suppressed at low temperatures by the
Boltzmann factor, with
  $\eta S_{\rm H} < 10^{-18}$ at $50~\K$ across all
hydrogen densities. In contrast, the quantum simulations
reveal sustained and efficient \chem{H_2} formation below
$100~\K$, as tunneling mitigates the classical Boltzmann
suppression. At low temperatures under
  the thermalization assumption (Figure \ref{fig:kmc},
  panel~a), where we can ignore hopping and desorption due
  to their high barriers, the two-H association becomes
  nearly barrierless, shifting the rate-limiting step to
  adsorption. As temperature rises, quantum-enhanced
  desorption begins to dominate, particularly under low
  hydrogen density conditions, leading to a decline in net
  formation. Remarkably, at $200~\K$ and
  $n(\mathrm{H}) = 10^2~\cm^{-3}$, the quantum \chem{H_2}
  formation efficiency even dips below the classical value,
  illustrating the nuanced and non-uniform impact of NQEs
  across parameter space.




  Under the adiabatic assumption (Figure
  \ref{fig:kmc}, panel~b), since $T_{\mathrm{gas}}$ (set at
  $600~\K$) is sufficiently high that NQEs can be neglected
  in environments such as PDRs, adsorption is treated as H
  atoms overcoming a potential energy barrier with kinetic
  energies consistent with $T_{\mathrm{gas}}$. This makes
  the adsorption step more kinetically favorable (i.e., a
  higher sticking probability) than under the thermalization
  assumption. At low dust temperatures as we discussed above
  H adsorption is the rate-limiting step in \chem{H_2}
  formation. When swithing from the thermalization
  assumption to the adiabatic assumption, the higher
  sticking probability greatly increases the formation refficiency.

  The KMC study on the \chem{MgSiO_3}
  surface (modeled as a 1D chain of $\sim 1000$ adsorption
  sites) reveals a qualitatively similar temperature-density
  dependence under the thermalization assumptions (Figure
  \ref{fig:kmc}, panel~c). However, the increase to
  $T_{\mathrm{gas}}\sim 600~\K$ under the adiabatic
  assumption results in only a slight change in the
  \chem{H_2} formation rates and $\eta S_{\rm H}$ compared
  to the thermalization assumption results (Figure
  \ref{fig:kmc}, panel~d). The entirely distinct
  dependencies on $T_{\mathrm{gas}}$ reflect the different
  chemical mechanism of \chem{H_2} formation on the graphene
  and \chem{MgSiO_3} surface. We find that the active
  barrier of H adsorption is zero and the rate-limiting step
  is the two-H association at low $T_{\mathrm{dust}}$ on the
  \chem{MgSiO_3} surface. We summarize the influence of
  temperature and atomic hydrogen density in Appendix
  \ref{sec:kmc} in detail.


\subsection{Chemisorbed hydrogen and gas-dust
    temperature decoupling}

The quantitative
framework presented in this work represents a significant
departure from traditional astrochemical modeling, 
providing a more rigorous physical solution 
that chemisorbed hydrogen atoms can form
\chem{H_2} with high efficiency, simultaneously resolving
the Boltzmann-factor suppression at low temperatures
($\sim 20~\K$) and the rapid desorption of physisorbed atoms
at higher temperatures ($\sim 200~\K$). Our
MLFF-enabled CPIHMC simulations demonstrate that NQEs
fundamentally reshape \textbf{the entire potential} energy
landscape, resulting in much lower activation free energies for all
elementary steps (adsorption, diffusion, and association) at low temperatures.
Negligible adsorption and association barrier eliminates the traditional dependency on
physisorption-mediated models (e.g., Langmuir-Hinshelwood
mechanisms) at low temperatures, which was largely a response to
the impenetrable classical barriers of chemisorption
($>0.5~\eV$) \citep{Vidali2013ChemRev, Cazaux2016SR}, and the need for the {\it ad hoc}
formation-rate multipliers traditionally used to reconcile
observations \citep{GriecoEtAl2023NA,
  FeldmannEtAl2023MNRAS}.

  As our results emphasize the importance
  of the initial adsorption step that typically act as
  the bottleneck of the catalysis efficiency, we consider
  two limiting assumptions for gas-dust temperature
  coupling: (1) the thermalization assumption for H
  adsorption (H atoms thermalize with the dust before
  adsorption, $T_{\mathrm{gas}} = T_{\mathrm{dust}}$), and
  (2) the adiabatic assumption for H adsorption (H atoms
  retain gas phase kinetic energy until after adsorption,
  $T_{\mathrm{gas}} \gg T_{\mathrm{dust}}$). These two
  limits bracket the plausible range of behavior in
  different astrophysical environments, dense, cold clouds
  favour the thermalization limit, while PDRs tend toward
  the adiabatic limit. By exploring both limits, we provide
  a more comprehensive picture of how gas-dust temperature
  decoupling may affect \chem{H_2} formation in different
  astrophysical environments. Importantly, our central
  conclusion remains robust across both assumptions: NQEs
  are essential for accurately describing \chem{H_2}
  formation on interstellar grain surfaces over a wide
  temperature range. We also find that switching from the
  thermalization assumption to the adiabatic assumption for
  H adsorption significantly alters the \chem{H_2} formation
  behavior on graphene, while leaving that on \chem{MgSiO_3}
  nearly unchanged. This difference arises from the distinct
  rate determining steps on the two surfaces: adsorption on
  graphene versus two-H association on \chem{MgSiO_3} due to
  the barrierless H adsorption.

\section{Discussion and Summary}
\label{sec:summary}

\subsection{Astrophysical applications}
\label{sec:astro-app}

Recent cosmological and galaxy-formation simulations
underscore the necessity of a consistent NQEs-based rate
prescription. Models adopting empirical formation rates
often require an efficiency factor significantly greater
than unity to explain the observed star-formation rates at
high redshifts, where dust abundance is low
\citep{FeldmannEtAl2023MNRAS, 2023MNRAS.519.3154H,
  2024A&A...691A.200R}. When the classical prescription is
replaced by an observation-calibrated rate that implicitly
includes NQEs, the critical metallicity for molecule
formation naturally drops from
$Z_{\mathrm{crit}} \sim 10^{-1} \, Z_\odot$ to the expected
$\lesssim 10^{-2} \, Z_\odot$
\citep{FeldmannEtAl2023MNRAS}. This shift enables
Population-II dust at $z \gtrsim 6$ to build molecular
clouds long before the temperature of the intergalactic
medium falls below $\sim 30~\K$. The same rate law naturally
reproduces the observed Kennicutt-Schmidt index
$N \approx 1.4$ without additional tuning and explains why
translucent clouds with $n(\chem{H}) \approx 300~\cm^{-3}$
and $T \approx 80~\K$ already harbour a significant
molecular fraction $f_{\chem{H_2}} \approx 0.1$
\citep{Habart2004AAP}. From Milky-Way dark clouds to the
first galaxies, NQEs thus provide a universal,
observationally-calibrated backbone for the chemistry that
sets the star-formation threshold across cosmic time.

We have also noticed that
several alternative astrophysical mechanisms may also enable
efficient molecular hydrogen production on grain surfaces,
even when the classical thermal pathway is suppressed. One
such possibility is a decoupling between the gas temperature
and the dust temperature. In environments with relatively
important photodissociation, photoionization, or shocks,
such as photodissociation regions (PDRs), the gas
temperature can rise to $T\sim 100~\K$ or even $\sim 600~\K$
(see e.g., the summaries in \citealt{Draine_book}).  Under
these conditions, the chemisorption rate increases
substantially, leading to significantly higher adsorption
and subsequent \chem{H_2} formation efficiency, even if the
dust grains themselves remain cold (as shown in
Figrue~\ref{fig:kmc}). This gas-dust temperature decoupling
thus provides a complementary channel for quantum-assisted
formation at warm gas temperatures.

Another promising mechanism involves turbulent motions in
the interstellar medium, which can produce large relative
(or ``slip'') velocities between dust grains and the
surrounding gas. In typical interstellar turbulence, the
expected slip velocities can reach $\sim 1-10~\km~\s^{-1}$
\citep[e.g.][]{2004ApJ...616..895Y}. Notably, the effective
potential barrier for adsorption is typically $\sim 0.1~\eV$
(Figure~\ref{fig:free_energy}; see also
\citealt{Horneker2006PRL, Zecho2002JCP}), which corresponds
to a kinetic energy equivalent to a hydrogen atom at
velocity $\sim 4~\km~\s^{-1}$. Therefore, when turbulence
imparts such velocities to hydrogen atoms impinging on a
dust grain, the atoms can directly overcome the adsorption
barrier without requiring thermal assistance or quantum
tunneling. This turbulence-induced pathway may therefore
contribute significantly to \chem{H_2} formation in
dynamically active regions, complementing the NQEs-dominated
channel in quiescent clouds. Both mechanisms, including
dust-gas decoupling and turbulent slips, highlight the
importance of considering non-thermal processes alongside
quantum effects in a complete astrochemical model.

\subsection{Future works}

In future works, the methodology developed in this study can
be directly applied to constrain the absolute \chem{H_2}
formation efficiency across the full spectrum of
interstellar environments. A key application lies in
predicting the non-thermal rotational and vibrational
excitation of newly formed \chem{H_2} molecules, which carry
away the $\sim 4.5~\eV$ binding energy released during
association. Observational signatures of this excess
excitations, which are detectable in infrared and
ultraviolet spectra, serve as direct probes of the formation
mechanism. Consistent and full-procedure quantum mechanical
simulations 
are essential for accurately calculating the energy
partition among translational, rotational, and vibrational
degrees of freedom, particularly in the low-temperature
regime ($T \lesssim 100~\K$) where NQEs dominate the
reaction dynamics. Such calculations would enable direct
comparison with upcoming high-resolution spectroscopic data
from facilities like JWST and the planned Origins Space
Telescope, providing a stringent test of our
quantum-tunneling model. 

  The multiscale simulation framework can also be
  applied to other relevant reactions on dust grains beyond
  traditional semi-quantum models, including the formation
  of water ice (\chem{H_2O}; see
  e.g. \citealt{2012ApJ...749...67O}) or more complex
  organic molecules via successive hydrogenation of CO, C,
  N, and O atoms. NQEs in these processes remain largely
  unexplored at dense-cloud temperatures but are likely
  equally decisive. Our path-integral treatment can
  incorporate heavier atoms (C, N, O) within the same
  formalism, enabling fully quantum studies across a broad
  range of astrochemical surface reactions.




\begin{acknowledgments}
  The authors acknowledge funding support from the National
  Natural Science Foundation of China (grant no. 92470114,
  no. 52273223, no. 12573067), School of Materials Science
  and Engineering at Peking University, and the AI for
  Science Institute, Beijing (AISI). The computing resource
  of this work was provided by the Bohrium Cloud Platform
  (https://bohrium.dp.tech), which was supported by DP
  Technology.
\end{acknowledgments}

\begin{contribution}
  L. W. and S. X. designed the project. X. Y. performed the
  whole calculations, reduced and analyzed the
  data. X. Y. were the primary writer of the manuscript. All
  the authors contributed to the discussion and the
  scientific interpretation of the results. L. W. and
  S. X. reviewed the manuscript.
\end{contribution}

\bibliography{manuscript}{}

@ARTICLE{2004ApJ...616..895Y,
       author = {{Yan}, Huirong and {Lazarian}, A. and {Draine}, B.~T.},
        title = "{Dust Dynamics in Compressible Magnetohydrodynamic Turbulence}",
      journal = {\apj},
         year = 2004,
        month = dec,
       volume = 616,
       number = 2,
        pages = {895-911},
          doi = {10.1086/425111},
archivePrefix = {arXiv},
       eprint = {astro-ph/0408173},
 primaryClass = {astro-ph},
       adsurl = {https://ui.adsabs.harvard.edu/abs/2004ApJ...616..895Y}
}

@article{MieEtAl2019FiC,
comment = RN27,
   author = {Andersen, Mie and Panosetti, Chiara and Reuter, Karsten},
   title = {A Practical Guide to Surface Kinetic Monte Carlo Simulations},
   journal = {Front. Chem.},
   volume = {Volume 7 - 2019},
   ISSN = {2296-2646},
   DOI = {10.3389/fchem.2019.00202},
   url = {https://www.frontiersin.org/journals/chemistry/articles/10.3389/fchem.2019.00202},
   year = {2019},
   type = {Journal Article}
}

@article{Cazaux2002ApJ,
comment = RN20,
   author = {Cazaux, S. and Tielens, A. G. G. M.},
   title = {Molecular Hydrogen Formation in the Interstellar Medium},
   journal = {\apj},
   volume = {575},
   number = {1},
   pages = {L29},
   ISSN = {0004-637X},
   DOI = {10.1086/342607},
   url = {https://dx.doi.org/10.1086/342607},
   year = {2002},
   type = {Journal Article}
}

@article{Cazaux2016SR,
comment = RN46,
   author = {Cazaux, Stéphanie and Boschman, Leon and
                  Rougeau, Nathalie and Reitsma, Geert and
                  Hoekstra, Ronnie and Teillet-Billy,
                  Dominique and Morisset, Sabine and Spaans,
                  Marco and Schlathoelter, Thomas},
   title = {The sequence to hydrogenate coronene cations: A
                  journey guided by magic numbers},
   journal = {Sci. Rep.},
   volume = {6},
   number = {1},
   pages = {19835},
   ISSN = {2045-2322},
   DOI = {10.1038/srep19835},
   url = {https://doi.org/10.1038/srep19835},
   year = {2016},
   type = {Journal Article}
}

@article{Chen2017ApJ,
   author = {Chen, X. H. and Li, Aigen and Zhang, Ke},
   title = {On Graphene in the Interstellar Medium},
   journal = {\apj},
   volume = {850},
   number = {1},
   pages = {104},
   ISSN = {0004-637X},
   DOI = {10.3847/1538-4357/aa93d5},
   url = {https://dx.doi.org/10.3847/1538-4357/aa93d5},
   year = {2017},
   type = {Journal Article}
}

@article{christensen2012mnras,
   author = {Christensen, Charlotte and Quinn, Thomas and Governato, Fabio and Stilp, Adrienne and Shen, Sijing and Wadsley, James},
   title = {Implementing molecular hydrogen in hydrodynamic simulations of galaxy formation},
   journal = {\mnras},
   volume = {425},
   number = {4},
   pages = {3058-3076},
   ISSN = {0035-8711},
   DOI = {10.1111/j.1365-2966.2012.21628.x},
   url = {https://doi.org/10.1111/j.1365-2966.2012.21628.x},
   year = {2012},
   type = {Journal Article}
}

@article{Den2000JCP,
   author = {Den Otter, W. K.},
   title = {Thermodynamic integration of the free energy along a reaction coordinate in Cartesian coordinates},
   journal = {\jcp},
   volume = {112},
   number = {17},
   pages = {7283-7292},
   ISSN = {0021-9606},
   DOI = {10.1063/1.481329},
   url = {https://dx.doi.org/10.1063/1.481329},
   year = {2000},
   type = {Journal Article}
}

@article{Draine2003ARAA,
comment = RN12,
   author = {Draine, B. T.},
   title = {Interstellar Dust Grains},
   journal = {\araa},
   volume = {41},
   number = {1},
   pages = {241-289},
   ISSN = {0066-4146},
   DOI = {10.1146/annurev.astro.41.011802.094840},
   url = {https://dx.doi.org/10.1146/annurev.astro.41.011802.094840},
   year = {2003},
   type = {Journal Article}
}

@article{FeldmannEtAl2023MNRAS,
comment = RN48,
   author = {Feldmann, Robert and Quataert, Eliot and
                  Faucher-Giguère, Claude-André and
                  Hopkins, Philip F and Çatmabacak, Onur and
                  Kereš, Dušan and Bassini, Luigi and
                  Bernardini, Mauro and Bullock, James S and
                  Cenci, Elia and Gensior, Jindra and Liang,
                  Lichen and Moreno, Jorge and Wetzel,
                  Andrew},
   title = {FIREbox: simulating galaxies at high dynamic
                  range in a cosmological volume},
   journal = {\mnras},
   volume = {522},
   number = {3},
   pages = {3831-3860},
   ISSN = {0035-8711},
   DOI = {10.1093/mnras/stad1205},
   url = {https://doi.org/10.1093/mnras/stad1205},
   year = {2023},
   type = {Journal Article}
}

@article{Feynman1948PI,
   author = {Feynman, R. P.},
   title = {Space-Time Approach to Non-Relativistic Quantum Mechanics},
   journal = {Rev. Mod. Phys.},
   volume = {20},
   number = {2},
   pages = {367-387},
   ISSN = {0034-6861},
   DOI = {10.1103/revmodphys.20.367},
   url = {https://dx.doi.org/10.1103/revmodphys.20.367},
   year = {1948},
   type = {Journal Article}
}

@article{GnedinKravtsovApJ2010,
comment = RN2,
   author = {Gnedin, Nickolay Y. and Kravtsov, Andrey V.},
   title = {ON THE KENNICUTT–SCHMIDT RELATION OF
                  LOW-METALLICITY HIGH-REDSHIFT GALAXIES},
   journal = {\apj},
   volume = {714},
   number = {1},
   pages = {287},
   ISSN = {0004-637X},
   DOI = {10.1088/0004-637X/714/1/287},
   url = {https://dx.doi.org/10.1088/0004-637X/714/1/287},
   year = {2010},
   type = {Journal Article}
}

@article{gnedin2009apj,
   author = {Gnedin, Nickolay Y. and Tassis, Konstantinos and Kravtsov, Andrey V.},
   title = {Modeling Molecular Hydrogen and Star Formation in Cosmological Simulations},
   journal = {\apj},
   volume = {697},
   pages = {55-67},
   ISSN = {0004-637X},
   DOI = {10.1088/0004-637x/697/1/55},
   url = {https://ui.adsabs.harvard.edu/abs/2009ApJ...697...55G},
   year = {2009},
   type = {Journal Article}
}

@article{GoldsmithLi2005ApJHINSA,
comment = RN43,
   author = {Goldsmith, P. F. and Li, D.},
   title = {H<scp>i</scp>Narrow Self‐Absorption in Dark Clouds: Correlations with Molecular Gas and Implications for Cloud Evolution and Star Formation},
   journal = {\apj},
   volume = {622},
   number = {2},
   pages = {938-958},
   ISSN = {0004-637X},
   DOI = {10.1086/428032},
   url = {https://dx.doi.org/10.1086/428032},
   year = {2005},
   type = {Journal Article}
}

@article{GriecoEtAl2023NA,
comment = RN47,
   author = {Grieco, Francesco and Theulé, Patrice and De Looze, Ilse and Dulieu, François},
   title = {Enhanced star formation through the high-temperature formation of H2 on carbonaceous dust grains},
   journal = {Nat. Astron.},
   volume = {7},
   number = {5},
   pages = {541-545},
   ISSN = {2397-3366},
   DOI = {10.1038/s41550-023-01902-4},
   url = {https://doi.org/10.1038/s41550-023-01902-4},
   year = {2023},
   type = {Journal Article}
}

@article{Habart2004AAP,
comment = RN44,
   author = {Habart, E. and Boulanger, F. and Verstraete,
                  L. and Walmsley, C. M. and Forêts,
                  G. Pineau Des},
   title = {Some empirical estimates of the H2formation rate
                  in photon-dominated regions},
   journal = {\aap},
   volume = {414},
   number = {2},
   pages = {531-544},
   ISSN = {0004-6361},
   DOI = {10.1051/0004-6361:20031659},
   url = {https://dx.doi.org/10.1051/0004-6361:20031659},
   year = {2004},
   type = {Journal Article}
}

@article{HanEtAl2022JPCL,
comment = RN22,
   author = {Han, Erxun and Fang, Wei and Stamatakis, Michail and Richardson, Jeremy O. and Chen, Ji},
   title = {Quantum Tunnelling Driven H2 Formation on Graphene},
   journal = {J. Phys. Chem. Lett.},
   volume = {13},
   number = {14},
   pages = {3173-3181},
   DOI = {10.1021/acs.jpclett.2c00520},
   url = {https://doi.org/10.1021/acs.jpclett.2c00520},
   year = {2022},
   type = {Journal Article}
}

@article{Henning2010araa,
   author = {Henning, Thomas},
   title = {Cosmic Silicates},
   journal = {\araa},
   volume = {48},
   number = {Volume 48, 2010},
   pages = {21-46},
   ISSN = {1545-4282},
   DOI = {https://doi.org/10.1146/annurev-astro-081309-130815},
   url = {https://www.annualreviews.org/content/journals/10.1146/annurev-astro-081309-130815},
   year = {2010},
   type = {Journal Article}
}

@article{Herbst2001review,
   author = {Herbst, Eric},
   title = {The chemistry of interstellar space},
   journal = {Chem. Soc. Rev.},
   volume = {30},
   number = {3},
   pages = {168-176},
   ISSN = {03060012
14604744},
   DOI = {10.1039/a909040a},
   year = {2001},
   type = {Journal Article}
}

@article{Hollenbach1971ApJ,
comment = RN19,
   author = {Hollenbach, David and Salpeter, E. E.},
   title = {Surface Recombination of Hydrogen Molecules},
   journal = {\apj},
   volume = {163},
   pages = {155},
   ISSN = {0004-637X},
   DOI = {10.1086/150754},
   url = {https://ui.adsabs.harvard.edu/abs/1971ApJ...163..155H},
   year = {1971},
   type = {Journal Article}
}

@article{Horneker2006PRL,
   author = {{Hornek{\ae}r}, L. and Rauls, E. and Xu, W. and Šljivančanin, Ž and Otero, R. and Stensgaard, I. and Lægsgaard, E. and Hammer, B. and Besenbacher, F.},
   title = {Clustering of Chemisorbed H(D) Atoms on the Graphite (0001) Surface due to Preferential Sticking},
   journal = {\prl},
   volume = {97},
   number = {18},
   pages = {186102},
   DOI = {10.1103/PhysRevLett.97.186102},
   url = {https://link.aps.org/doi/10.1103/PhysRevLett.97.186102},
   year = {2006},
   type = {Journal Article}
}

@article{MatProj2013,
comment = RN38,
   author = {Jain, Anubhav and Ong, Shyue Ping and Hautier, Geoffroy and Chen, Wei and Richards, William Davidson and Dacek, Stephen and Cholia, Shreyas and Gunter, Dan and Skinner, David and Ceder, Gerbrand and Persson, Kristin A.},
   title = {Commentary: The Materials Project: A materials genome approach to accelerating materials innovation},
   journal = {APL Mater.},
   volume = {1},
   number = {1},
   ISSN = {2166-532X},
   DOI = {10.1063/1.4812323},
   url = {https://doi.org/10.1063/1.4812323},
   year = {2013},
   type = {Journal Article}
}

@article{Jin2023JCTC,
   author = {Jin, Bin and Hu, Taiping and Yu, Kuang and Xu, Shenzhen},
   title = {Constrained Hybrid Monte Carlo Sampling Made Simple for Chemical Reaction Simulations},
   journal = {J. Chem. Theory Comput.},
   volume = {19},
   number = {20},
   pages = {7343-7357},
   ISSN = {1549-9618},
   DOI = {10.1021/acs.jctc.3c00571},
   url = {https://doi.org/10.1021/acs.jctc.3c00571},
   year = {2023},
   type = {Journal Article}
}

@article{Kohn1965DFT,
   author = {Kohn, W. and Sham, L. J.},
   title = {Self-Consistent Equations Including Exchange and Correlation Effects},
   journal = {Phys. Rev.},
   volume = {140},
   number = {4A},
   pages = {A1133-A1138},
   ISSN = {0031-899X},
   DOI = {10.1103/physrev.140.a1133},
   url = {https://dx.doi.org/10.1103/physrev.140.a1133},
   year = {1965},
   type = {Journal Article}
}

@article{Lu2021CPCdpmd,
   author = {Lu, Denghui and Wang, Han and Chen, Mohan and Lin, Lin and Car, Roberto and E, Weinan and Jia, Weile and Zhang, Linfeng},
   title = {86 PFLOPS Deep Potential Molecular Dynamics simulation of 100 million atoms with ab initio accuracy},
   journal = {Comput. Phys. Commun.},
   volume = {259},
   pages = {107624},
   ISSN = {0010-4655},
   DOI = {https://doi.org/10.1016/j.cpc.2020.107624},
   url = {https://www.sciencedirect.com/science/article/pii/S001046552030299X},
   year = {2021},
   type = {Journal Article}
}

@article{Molster2002AAa,
   author = {Molster, F. J. and Waters, L. B. F. M. and Tielens, A. G. G. M.},
   title = {Crystalline silicate dust around evolved stars},
   journal = {\aap},
   volume = {382},
   number = {1},
   pages = {222-240},
   ISSN = {0004-6361},
   DOI = {10.1051/0004-6361:20011551},
   url = {https://dx.doi.org/10.1051/0004-6361:20011551},
   year = {2002},
   type = {Journal Article}
}

@article{Molster2002AAb,
   author = {Molster, F. J. and Waters, L. B. F. M. and Tielens, A. G. G. M. and Koike, C. and Chihara, H.},
   title = {Crystalline silicate dust around evolved stars*},
   journal = {\aap},
   volume = {382},
   number = {1},
   pages = {241-255},
   url = {https://doi.org/10.1051/0004-6361:20011552},
   year = {2002},
   type = {Journal Article}
}

@article{shull1982araa,
   author = {Shull, J. M. and Beckwith, S.},
   title = {Interstellar molecular hydrogen},
   journal = {\araa},
   volume = {20},
   pages = {163-190},
   ISSN = {0066-4146},
   DOI = {10.1146/annurev.aa.20.090182.001115},
   url = {https://ui.adsabs.harvard.edu/abs/1982ARA&A..20..163S},
   year = {1982},
   type = {Journal Article}
}

@article{Sprik1998JCP,
   author = {Sprik, M. and Ciccotti, G.},
   title = {Free energy from constrained molecular dynamics},
   journal = {\jcp},
   volume = {109},
   number = {18},
   pages = {7737-7744},
   ISSN = {0021-9606},
   DOI = {10.1063/1.477419},
   url = {<Go to ISI>://WOS:000076663100009},
   year = {1998},
   type = {Journal Article}
}

@article{Sun2025NC,
   author = {Sun, Menglin and Jin, Bin and Yang, Xiaolong and Xu, Shenzhen},
   title = {Probing nuclear quantum effects in electrocatalysis via a machine-learning enhanced grand canonical constant potential approach},
   journal = {Nat. Commun.},
   volume = {16},
   number = {1},
   pages = {3600},
   ISSN = {2041-1723},
   DOI = {10.1038/s41467-025-58871-7},
   url = {https://doi.org/10.1038/s41467-025-58871-7},
   year = {2025},
   type = {Journal Article}
}

@article{2026Guobing,
   author = {Zhou, Guobing and Hu, Taiping and Jin, Bin and Fu, Fangjia and Wang, Haoyu and Zhang, Hao and Yang, Zhen and Xu, Shenzhen},
   title = {Atomic Resolution of Solid–Electrolyte Interphase Formation via Off-Lattice On-the-Fly Kinetic Monte Carlo},
   journal = {Journal of the American Chemical Society},
   volume = {148},
   number = {1},
   pages = {2059-2070},
   ISSN = {0002-7863},
   DOI = {10.1021/jacs.5c21439},
   url = {https://doi.org/10.1021/jacs.5c21439},
   year = {2026},
   type = {Journal Article}
}

@article{Tong2024JPCC,
comment = RN25,
   author = {Tong, Yangwu and Yang, Yong},
   title = {Hydrogen Diffusion on Graphene Surface: The Effects of Neighboring Adsorbate and Quantum Tunneling},
   journal = {J. Phys. Chem. C},
   volume = {128},
   number = {2},
   pages = {840-849},
   ISSN = {1932-7447},
   DOI = {10.1021/acs.jpcc.3c05315},
   url = {https://doi.org/10.1021/acs.jpcc.3c05315},
   year = {2024},
   type = {Journal Article}
}

@article{Tong2024CPL,
comment = RN26,
   author = {Tong, Yangwu and Yang, Yong},
   title = {Quantum Tunneling Enhanced Hydrogen Desorption from Graphene Surface: Atomic versus Molecular Mechanism},
   journal = {Chin. Phys. Lett.},
   volume = {41},
   number = {8},
   pages = {086801},
   ISSN = {0256-307X},
   DOI = {10.1088/0256-307X/41/8/086801},
   url = {https://dx.doi.org/10.1088/0256-307X/41/8/086801},
   year = {2024},
   type = {Journal Article}
}

@article{Vidali2013ChemRev,
comment = RN21,
   author = {Vidali, G.},
   title = {H2 formation on interstellar grains},
   journal = {Chem. Rev.},
   volume = {113},
   number = {12},
   pages = {8762-82},
   ISSN = {1520-6890 (Electronic)
0009-2665 (Linking)},
   DOI = {10.1021/cr400156b},
   url = {https://www.ncbi.nlm.nih.gov/pubmed/24160443},
   year = {2013},
   type = {Journal Article}
}

@article{WangEtAL2018DPMDKit,
comment = RN28,
   author = {Wang, H. and Zhang, L. F. and Han, J. Q. and E, W. N.},
   title = {DeePMD-kit: A deep learning package for many-body potential energy representation and molecular dynamics},
   journal = {Comput. Phys. Commun.},
   volume = {228},
   pages = {178-184},
   ISSN = {0010-4655},
   DOI = {10.1016/j.cpc.2018.03.016},
   url = {<Go to ISI>://WOS:000434000900019},
   year = {2018},
   type = {Journal Article}
}

@article{Zecho2002JCP,
comment = RN49,
   author = {Zecho, Thomas and Güttler, Andreas and Sha, Xianwei and Jackson, Bret and Küppers, Jürgen},
   title = {Adsorption of hydrogen and deuterium atoms on the (0001) graphite surface},
   journal = {\jcp},
   volume = {117},
   number = {18},
   pages = {8486-8492},
   ISSN = {0021-9606},
   DOI = {10.1063/1.1511729},
   url = {https://doi.org/10.1063/1.1511729},
   year = {2002},
   type = {Journal Article}
}

@article{ZhangEtAl2020DPGEN,
comment = RN36,
   author = {Zhang, Y. Z. and Wang, H. D. and Chen, W. J. and Zeng, J. Z. and Zhang, L. F. and Wang, H. and Weinan, E.},
   title = {DP-GEN: A concurrent learning platform for the generation of reliable deep learning based potential energy models},
   journal = {Comput. Phys. Commun.},
   volume = {253},
   pages = {11},
   ISSN = {0010-4655},
   DOI = {10.1016/j.cpc.2020.107206},
   url = {<Go to ISI>://WOS:000537843600017},
   year = {2020},
   type = {Journal Article}
}

@book{Tuckerman2023book,
    author = {Tuckerman, Mark E.},
    title = {Statistical Mechanics: Theory and Molecular Simulation},
    publisher = {Oxford University Press},
    year = {2023},
    month = {08},
    isbn = {9780198825562},
    doi = {10.1093/oso/9780198825562.001.0001},
    url = {https://doi.org/10.1093/oso/9780198825562.001.0001},
}

@article{Bloechl1994PRBpaw,
   author = {Bloechl, P. E.},
   title = {Projector augmented-wave method},
   journal = {\prb},
   volume = {50},
   number = {24},
   pages = {17953-17979},
   ISSN = {0163-1829},
   DOI = {10.1103/physrevb.50.17953},
   url = {https://dx.doi.org/10.1103/physrevb.50.17953},
   year = {1994},
   type = {Journal Article}
}

@article{Dobson2002AJC,
   author = {Dobson, John F. and McLennan, Keith and Rubio, Angel and Wang, Jun and Gould, Tim and Le, Hung M. and Dinte, Bradley P.},
   title = {Prediction of Dispersion Forces: Is There a Problem?},
   journal = {Aust. J. Chem.},
   volume = {54},
   number = {8},
   pages = {513-527},
   DOI = {https://doi.org/10.1071/CH01052},
   url = {https://www.publish.csiro.au/paper/CH01052},
   year = {2002},
   type = {Journal Article}
}

@article{Grimme2010JCPD3,
   author = {Grimme, Stefan and Antony, Jens and Ehrlich, Stephan and Krieg, Helge},
   title = {A consistent and accurate <i>ab initio</i> parametrization of density functional dispersion correction (DFT-D) for the 94 elements H-Pu},
   journal = {\jcp},
   volume = {132},
   number = {15},
   pages = {154104},
   ISSN = {0021-9606},
   DOI = {10.1063/1.3382344},
   url = {https://dx.doi.org/10.1063/1.3382344},
   year = {2010},
   type = {Journal Article}
}

@article{Kresse1996CMSvasp,
   author = {Kresse, G. and Furthmuller, J.},
   title = {Efficiency of ab-initio total energy calculations for metals and semiconductors using a plane-wave basis set},
   journal = {Comput. Mater. Sci.},
   volume = {6},
   number = {1},
   pages = {15-50},
   ISSN = {0927-0256},
   DOI = {10.1016/0927-0256(96)00008-0},
   url = {<Go to ISI>://WOS:A1996VF38900003},
   year = {1996},
   type = {Journal Article}
}

@article{Kresse1996PRBvasp,
   author = {Kresse, G. and Furthmüller, J.},
   title = {Efficient iterative schemes for<i>ab initio</i>total-energy calculations using a plane-wave basis set},
   journal = {\prb},
   volume = {54},
   number = {16},
   pages = {11169-11186},
   ISSN = {0163-1829},
   DOI = {10.1103/physrevb.54.11169},
   url = {https://dx.doi.org/10.1103/physrevb.54.11169},
   year = {1996},
   type = {Journal Article}
}

@article{Mehlig1992PRB:hmc,
   author = {Mehlig, B. and Heermann, D. W. and Forrest, B. M.},
   title = {HYBRID MONTE-CARLO METHOD FOR CONDENSED-MATTER SYSTEMS},
   journal = {\prb},
   volume = {45},
   number = {2},
   pages = {679-685},
   ISSN = {0163-1829},
   DOI = {10.1103/PhysRevB.45.679},
   url = {<Go to ISI>://WOS:A1992HA07400016},
   year = {1992},
   type = {Journal Article}
}

@article{Metropolis1953JCP,
   author = {Metropolis, N. and Rosenbluth, Arianna W. and Rosenbluth, Marshall N. and Teller, A. H. and Teller, Edward},
   title = {Equation of state calculations by fast computing machines},
   journal = {\jcp},
   volume = {21},
   pages = {1087-1092},
   year = {1953},
   type = {Journal Article}
}

@article{Perdew1996PRLpbe,
   author = {Perdew, John P. and Burke, Kieron and Ernzerhof, Matthias},
   title = {Generalized Gradient Approximation Made Simple [Phys. Rev. Lett. 77, 3865 (1996)]},
   journal = {\prl},
   volume = {78},
   number = {7},
   pages = {1396-1396},
   ISSN = {0031-9007},
   DOI = {10.1103/physrevlett.78.1396},
   url = {https://dx.doi.org/10.1103/physrevlett.78.1396},
   year = {1997},
   type = {Journal Article}
}

@article{Thompson2022CPClammps,
   author = {Thompson, A. P. and Aktulga, H. M. and Berger, R. and Bolintineanu, D. S. and Brown, W. M. and Crozier, P. S. and Veld, P. J. I. and Kohlmeyer, A. and Moore, S. G. and Nguyen, T. D. and Shan, R. and Stevens, M. J. and Tranchida, J. and Trott, C. and Plimpton, S. J.},
   title = {LAMMPS-a flexible simulation tool for particle-based materials modeling at the atomic, meso, and continuum scales},
   journal = {Comput. Phys. Commun.},
   volume = {271},
   ISSN = {0010-4655},
   DOI = {10.1016/j.cpc.2021.108171},
   url = {<Go to ISI>://WOS:000720461800009},
   year = {2022},
   type = {Journal Article}
}

@article{Zhang2019PRM,
   author = {Zhang, Linfeng and Lin, De-Ye and Wang, Han and
                  Car, Roberto and E, Weinan},
   title = {Active learning of uniformly accurate
                  interatomic potentials for materials
                  simulation},
   journal = {Phys. Rev. Mater.},
   volume = {3},
   number = {2},
   pages = {023804},
   DOI = {10.1103/PhysRevMaterials.3.023804},
   url = {https://link.aps.org/doi/10.1103/PhysRevMaterials.3.023804},
   year = {2019},
   type = {Journal Article}
}

@BOOK{Draine_book,
  author = {{Draine}, B.~T.},
  title = "{Physics of the Interstellar and Intergalactic
                  Medium}",
  booktitle = {Physics of the Interstellar and Intergalactic
                  Medium by Bruce T.~Draine.~Princeton
                  University Press, 2011.~ISBN:
                  978-0-691-12214-4},
  publisher = "Princeton University Press",
  year = 2011,
}

@ARTICLE{Goldsmith2025ApJ,
       author = {{Goldsmith}, Paul F. and {Wang}, Shengzhe
                  and {Wang}, Xin and {Skalidis}, Raphael
                  and {Fuller}, Gary A. and {Li}, Di and
                  {Tsai}, Chao-Wei and {Wang}, Lile and
                  {Quan}, Donghui},
        title = "{The H$_{2}$ Glow of a Quiescent Molecular
                  Cloud Observed with JWST}",
      journal = {\apjl},
         year = 2025,
        month = may,
       volume = {985},
       number = {1},
          eid = {L4},
        pages = {L4},
          doi = {10.3847/2041-8213/adcf9c},
archivePrefix = {arXiv},
       eprint = {2504.20164},
 primaryClass = {astro-ph.GA},
       adsurl = {https://ui.adsabs.harvard.edu/abs/2025ApJ...985L...4G}
}

@article{2010Kalv,
   author = {Kalv{\-a}ns, J. and Shmeld, I.},
   title = {Subsurface chemistry of mantles of interstellar dust grains in dark molecular cores},
   journal = {\aap},
   volume = {521},
   pages = {A37},
   url = {https://doi.org/10.1051/0004-6361/201014190},
   year = {2010},
   type = {Journal Article}
}

@ARTICLE{Wakelam2017MolAs,
       author = {{Wakelam}, Valentine and {Bron}, Emeric and
                  {Cazaux}, Stephanie and {Dulieu}, Francois
                  and {Gry}, C{\'e}cile and {Guillard},
                  Pierre and {Habart}, Emilie and
                  {Hornek{\ae}r}, Liv and {Morisset}, Sabine
                  and {Nyman}, Gunnar and et al.},
        title = "{H$_{2}$ formation on interstellar dust
                  grains: The viewpoints of theory,
                  experiments, models and observations}",
      journal = {Mol. Astrophys.},
         year = 2017,
        month = dec,
       volume = {9},
        pages = {1-36},
          doi = {10.1016/j.molap.2017.11.001},
archivePrefix = {arXiv},
       eprint = {1711.10568},
 primaryClass = {astro-ph.GA},
       adsurl = {https://ui.adsabs.harvard.edu/abs/2017MolAs...9....1W}
}

@ARTICLE{2005A&A...434..599C,
       author = {{Chang}, Q. and {Cuppen}, H.~M. and {Herbst}, E.},
        title = "{Continuous-time random-walk simulation of
                  H$_{2}$ formation on interstellar grains}",
      journal = {\aap},
         year = 2005,
        month = may,
       volume = {434},
       number = {2},
        pages = {599-611},
          doi = {10.1051/0004-6361:20041842},
       adsurl = {https://ui.adsabs.harvard.edu/abs/2005A&A...434..599C}
}

@article{2005Cuppen,
   author = {Cuppen, H. M. and Herbst, Eric},
   title = {Monte Carlo simulations of H<SUB>2</SUB> formation on grains of varying surface roughness},
   journal = {Monthly Notices of the Royal Astronomical Society},
   volume = {361},
   pages = {565-576},
   ISSN = {0035-8711},
   DOI = {10.1111/j.1365-2966.2005.09189.x},
   url = {https://ui.adsabs.harvard.edu/abs/2005MNRAS.361..565C},
   year = {2005},
   type = {Journal Article}
}

@article{2006Cuppen,
   author = {Cuppen, H. M. and Morata, O. and Herbst, Eric},
   title = {Monte Carlo simulations of H<SUB>2</SUB> formation on stochastically heated grains},
   journal = {Monthly Notices of the Royal Astronomical Society},
   volume = {367},
   pages = {1757-1765},
   ISSN = {0035-8711},
   DOI = {10.1111/j.1365-2966.2006.10079.x},
   url = {https://ui.adsabs.harvard.edu/abs/2006MNRAS.367.1757C},
   year = {2006},
   type = {Journal Article}
}

@article{2014Rimola,
   author = {Navarro-Ruiz, Javier and Sodupe, Mariona and Ugliengo, Piero and Rimola, Albert},
   title = {Interstellar H adsorption and H2 formation on the crystalline (010) forsterite surface: a B3LYP-D2* periodic study},
   journal = {Physical Chemistry Chemical Physics},
   volume = {16},
   number = {33},
   pages = {17447-17457},
   ISSN = {1463-9076},
   DOI = {10.1039/C4CP00819G},
   url = {http://dx.doi.org/10.1039/C4CP00819G},
   year = {2014},
   type = {Journal Article}
}

@ARTICLE{2015Rimola,
       author = {{Navarro-Ruiz}, Javier and {Mart{\'\i}nez-Gonz{\'a}lez}, Jos{\'e} {\'A}ngel and {Sodupe}, Mariona and {Ugliengo}, Piero and {Rimola}, Albert},
        title = "{Relevance of silicate surface morphology in interstellar H$_{2}$ formation. Insights from quantum chemical calculations}",
      journal = {\mnras},
         year = 2015,
        month = oct,
       volume = {453},
       number = {1},
        pages = {914-924},
          doi = {10.1093/mnras/stv1628},
       adsurl = {https://ui.adsabs.harvard.edu/abs/2015MNRAS.453..914N}
}

@article{2016Rimola,
   author = {Navarro-Ruiz, J. and Ugliengo, P. and Sodupe, M. and Rimola, A.},
   title = {Does Fe2+ in olivine-based interstellar grains play any role in the formation of H2? Atomistic insights from DFT periodic simulations},
   journal = {Chemical Communications},
   volume = {52},
   number = {42},
   pages = {6873-6876},
   ISSN = {1359-7345},
   DOI = {10.1039/C6CC02313D},
   url = {http://dx.doi.org/10.1039/C6CC02313D},
   year = {2016},
   type = {Journal Article}
}

@article{2013Kerkeni,
   author = {Kerkeni, Boutheïna and Bromley, Stefan T.},
   title = {Competing mechanisms of catalytic H<SUB>2</SUB> formation and dissociation on ultrasmall silicate nanocluster dust grains},
   journal = {Monthly Notices of the Royal Astronomical Society},
   volume = {435},
   pages = {1486-1492},
   ISSN = {0035-8711},
   DOI = {10.1093/mnras/stt1389},
   url = {https://ui.adsabs.harvard.edu/abs/2013MNRAS.435.1486K},
   year = {2013},
   type = {Journal Article}
}

@article{2015Kerkeni,
   author = {Oueslati, Ichraf and Kerkeni, Boutheïna and Bromley, Stefan T.},
   title = {Trends in the adsorption and reactivity of hydrogen on magnesium silicate nanoclusters},
   journal = {Physical Chemistry Chemical Physics},
   volume = {17},
   number = {14},
   pages = {8951-8963},
   ISSN = {1463-9076},
   DOI = {10.1039/C4CP05128A},
   url = {http://dx.doi.org/10.1039/C4CP05128A},
   year = {2015},
   type = {Journal Article}
}

@article{2017Kerkeni,
   author = {Kerkeni, Boutheïna and Bacchus-Montabonel, Marie-Christine and Bromley, Stefan T.},
   title = {How hydroxylation affects hydrogen adsorption and formation on nanosilicates},
   journal = {Molecular Astrophysics},
   volume = {7},
   pages = {1-8},
   ISSN = {2405-6758},
   DOI = {https://doi.org/10.1016/j.molap.2017.04.001},
   url = {https://www.sciencedirect.com/science/article/pii/S2405675816300434},
   year = {2017},
   type = {Journal Article}
}

@article{2019Kerkeni,
   author = {Kerkeni, Boutheïna and Bacchus-Montabonel, Marie-Christine and Shan, Xiao and Bromley, Stefan T.},
   title = {Understanding H2 Formation on Hydroxylated Pyroxene Nanoclusters: Ab Initio Study of the Reaction Energetics and Kinetics},
   journal = {The Journal of Physical Chemistry A},
   volume = {123},
   number = {43},
   pages = {9282-9291},
   ISSN = {1089-5639},
   DOI = {10.1021/acs.jpca.9b06713},
   url = {https://doi.org/10.1021/acs.jpca.9b06713},
   year = {2019},
   type = {Journal Article}
}

@article{2022Kerkeni,
   author = {Kerkeni, Boutheïna and Boukallaba, Malek and Hechmi, Mariem and Duflot, Denis and Toubin, Céline},
   title = {QM/MM Study of the H2 Formation on the Surface of a Water Ice Grain Doped With Formaldehyde: Molecular Dynamics and Reaction Kinetics},
   journal = {Frontiers in Astronomy and Space Sciences},
   volume = {Volume 9 - 2022},
   ISSN = {2296-987X},
   DOI = {10.3389/fspas.2022.807649},
   url = {https://www.frontiersin.org/journals/astronomy-and-space-sciences/articles/10.3389/fspas.2022.807649},
   year = {2022},
   type = {Journal Article}
}

@article{2025Kerkeni,
   author = {Mennella, V. and Suhasaria, T. and Kerkeni, B. and Ouerfelli, G.},
   title = {Hydroxylated Mg-rich Amorphous Silicates as Catalysts for Molecular Hydrogen Formation in the Interstellar Medium},
   journal = {The Astrophysical Journal},
   volume = {987},
   number = {2},
   pages = {207},
   ISSN = {0004-637X},
   DOI = {10.3847/1538-4357/ade2d3},
   url = {https://doi.org/10.3847/1538-4357/ade2d3},
   year = {2025},
   type = {Journal Article}
}

@article{1999Takahashi,
   author = {Takahashi, Junko and Masuda, Koichi and Nagaoka, Masataka},
   title = {The formation mechanism of molecular hydrogen on icy mantles of interstellar dust},
   journal = {Monthly Notices of the Royal Astronomical Society},
   volume = {306},
   number = {1},
   pages = {22-30},
   ISSN = {0035-8711},
   DOI = {10.1046/j.1365-8711.1999.02480.x},
   url = {https://doi.org/10.1046/j.1365-8711.1999.02480.x},
   year = {1999},
   type = {Journal Article}
}

@article{2013Watanabe,
   author = {Hama, Tetsuya and Watanabe, Naoki},
   title = {Surface Processes on Interstellar Amorphous Solid Water: Adsorption, Diffusion, Tunneling Reactions, and Nuclear-Spin Conversion},
   journal = {Chemical Reviews},
   volume = {113},
   number = {12},
   pages = {8783-8839},
   ISSN = {0009-2665},
   DOI = {10.1021/cr4000978},
   url = {https://doi.org/10.1021/cr4000978},
   year = {2013},
   type = {Journal Article}
}

@ARTICLE{2023MNRAS.519.3154H,
       author = {{Hopkins}, Philip F. and {Wetzel}, Andrew
                  and {Wheeler}, Coral and {Sanderson},
                  Robyn and {Grudi{\'c}}, Michael Y. and
                  {Sameie}, Omid and {Boylan-Kolchin},
                  Michael and {Orr}, Matthew and {Ma},
                  Xiangcheng and {Faucher-Gigu{\`e}re},
                  Claude-Andr{\'e} and {Kere{\v{s}}},
                  Du{\v{s}}an and {Quataert}, Eliot and
                  {Su}, Kung-Yi and {Moreno}, Jorge and
                  {Feldmann}, Robert and {Bullock}, James
                  S. and {Loebman}, Sarah R. and
                  {Angl{\'e}s-Alc{\'a}zar}, Daniel and
                  {Stern}, Jonathan and {Necib}, Lina and
                  {Choban}, Caleb R. and {Hayward},
                  Christopher C.},
        title = "{FIRE-3: updated stellar evolution models,
                  yields, and microphysics and fitting
                  functions for applications in galaxy
                  simulations}",
      journal = {\mnras},
         year = 2023,
        month = feb,
       volume = {519},
       number = {2},
        pages = {3154-3181},
          doi = {10.1093/mnras/stac3489},
archivePrefix = {arXiv},
       eprint = {2203.00040},
 primaryClass = {astro-ph.GA},
       adsurl = {https://ui.adsabs.harvard.edu/abs/2023MNRAS.519.3154H}
}

@ARTICLE{2024A&A...691A.200R,
       author = {{Ragone-Figueroa}, Cinthia and {Granato},
                  Gian Luigi and {Parente}, Massimiliano and
                  {Murante}, Giuseppe and {Valentini},
                  Milena and {Borgani}, Stefano and {Maio},
                  Umberto},
        title = "{Intertwined formation of H$_{2}$, dust,
                  and stars in cosmological simulations}",
      journal = {\aap},
         year = 2024,
        month = nov,
       volume = {691},
          eid = {A200},
        pages = {A200},
          doi = {10.1051/0004-6361/202451344},
archivePrefix = {arXiv},
       eprint = {2407.06269},
 primaryClass = {astro-ph.GA},
       adsurl = {https://ui.adsabs.harvard.edu/abs/2024A&A...691A.200R}
}

@ARTICLE{2012ApJ...749...67O,
       author = {{Oba}, Y. and {Watanabe}, N. and {Hama},
                  T. and {Kuwahata}, K. and {Hidaka}, H. and
                  {Kouchi}, A.},
        title = "{Water Formation through a Quantum
                  Tunneling Surface Reaction, OH + H$_{2}$,
                  at 10 K}",
      journal = {\apj},
         year = 2012,
        month = apr,
       volume = {749},
       number = {1},
          eid = {67},
        pages = {67},
          doi = {10.1088/0004-637X/749/1/67},
archivePrefix = {arXiv},
       eprint = {1202.1035},
 primaryClass = {astro-ph.GA},
       adsurl = {https://ui.adsabs.harvard.edu/abs/2012ApJ...749...67O}
}

@article{2010_tunnel_H2formation,
author = {Goumans, Theodorus P. M. and Kästner, Johannes},
title = {Hydrogen-Atom Tunneling Could Contribute to H2 Formation in Space},
journal = {Angewandte Chemie International Edition},
volume = {49},
number = {40},
pages = {7350-7352},
doi = {https://doi.org/10.1002/anie.201001311},
url = {https://onlinelibrary.wiley.com/doi/abs/10.1002/anie.201001311},
eprint = {https://onlinelibrary.wiley.com/doi/pdf/10.1002/anie.201001311},
year = {2010}
}

@article{1990Peter,
   author = {Hänggi, Peter and Talkner, Peter and Borkovec, Michal},
   title = {Reaction-rate theory: fifty years after Kramers},
   journal = {Reviews of Modern Physics},
   volume = {62},
   number = {2},
   pages = {251-341},
   note = {RMP},
   DOI = {10.1103/RevModPhys.62.251},
   url = {https://link.aps.org/doi/10.1103/RevModPhys.62.251},
   year = {1990},
   type = {Journal Article}
}

@article{1984Grabert,
   author = {Grabert, Hermann and Weiss, Ulrich and Hanggi, Peter},
   title = {Quantum Tunneling in Dissipative Systems at Finite Temperatures},
   journal = {Physical Review Letters},
   volume = {52},
   number = {25},
   pages = {2193-2196},
   note = {PRL},
   DOI = {10.1103/PhysRevLett.52.2193},
   url = {https://link.aps.org/doi/10.1103/PhysRevLett.52.2193},
   year = {1984},
   type = {Journal Article}
}

@article{2017Gran,
   author = {Granja-DelRío, Alejandra and Alonso, Julio A. and López, María J.},
   title = {Competition between Palladium Clusters and Hydrogen to Saturate Graphene Vacancies},
   journal = {The Journal of Physical Chemistry C},
   volume = {121},
   number = {20},
   pages = {10843-10850},
   ISSN = {1932-7447},
   DOI = {10.1021/acs.jpcc.6b12018},
   url = {https://doi.org/10.1021/acs.jpcc.6b12018},
   year = {2017},
   type = {Journal Article}
}

@article{2016Meisner,
   author = {Meisner, Jan and Kästner, Johannes},
   title = {Atom Tunneling in Chemistry},
   journal = {Angewandte Chemie International Edition},
   volume = {55},
   number = {18},
   pages = {5400-5413},
   ISSN = {1433-7851},
   DOI = {https://doi.org/10.1002/anie.201511028},
   url = {https://onlinelibrary.wiley.com/doi/abs/10.1002/anie.201511028},
   year = {2016},
   type = {Journal Article}
}

@article{2017Senevirathne,
   author = {Senevirathne, Bethmini and Andersson, Stefan and Dulieu, Francois and Nyman, Gunnar},
   title = {Hydrogen atom mobility, kinetic isotope effects and tunneling on interstellar ices (Ih and ASW)},
   journal = {Molecular Astrophysics},
   volume = {6},
   pages = {59-69},
   ISSN = {2405-6758},
   DOI = {https://doi.org/10.1016/j.molap.2017.01.005},
   url = {https://www.sciencedirect.com/science/article/pii/S2405675816300318},
   year = {2017},
   type = {Journal Article}
}

@phdthesis{2021Enrique-Romero,
   author = {Enrique-Romero, Joan},
   title = {Surface chemistry of molecules of astrophysical interest : theory and simulations
Chimie de surface d'espèces d'intérêt astrophysique : théorie et simulations},
   school = {Université Grenoble Alpes en cotutelle avec Universitat autònoma de Barcelona},
   url = {https://theses.hal.science/tel-03579109},
   year = {2021},
}

@article{1987Ando,
   author = {Ando, Yuji and Itoh, Tomohiro},
   title = {Calculation of transmission tunneling current across arbitrary potential barriers},
   journal = {Journal of Applied Physics},
   volume = {61},
   number = {4},
   pages = {1497-1502},
   ISSN = {0021-8979},
   DOI = {10.1063/1.338082},
   url = {https://doi.org/10.1063/1.338082},
   year = {1987},
   type = {Journal Article}
}

@article{2014Congiu,
   author = {Congiu, Emanuele and Minissale, Marco and Baouche, Saoud and Chaabouni, Henda and Moudens, Audrey and Cazaux, Stephanie and Manicò, Giulio and Pirronello, Valerio and Dulieu, François},
   title = {Efficient diffusive mechanisms of O atoms at very low temperatures on surfaces of astrophysical interest},
   journal = {Faraday Discussions},
   volume = {168},
   number = {0},
   pages = {151-166},
   ISSN = {1359-6640},
   DOI = {10.1039/C4FD00002A},
   url = {http://dx.doi.org/10.1039/C4FD00002A},
   year = {2014},
   type = {Journal Article}
}

@article{2010Goumans,
   author = {Goumans, Theodorus P. M. and Kästner, Johannes},
   title = {Hydrogen-Atom Tunneling Could Contribute to H2 Formation in Space},
   journal = {Angewandte Chemie International Edition},
   volume = {49},
   number = {40},
   pages = {7350-7352},
   ISSN = {1433-7851},
   DOI = {https://doi.org/10.1002/anie.201001311},
   url = {https://onlinelibrary.wiley.com/doi/abs/10.1002/anie.201001311},
   year = {2010},
   type = {Journal Article}
}

@article{2016Lamberts,
   author = {Lamberts, Thanja and Samanta, Pradipta Kumar and Koehn, Andreas and Kästner, Johannes},
   title = {Quantum tunneling during interstellar surface-catalyzed formation of water: the reaction H + H<sub>2</sub>O<sub>2</sub> → H<sub>2</sub>O + OH},
   journal = {Physical Chemistry Chemical Physics},
   volume = {18},
   number = {48},
   pages = {33021-33030},
   ISSN = {1463-9076},
   DOI = {10.1039/c6cp06457d},
   url = {https://dx.doi.org/10.1039/c6cp06457d},
   year = {2016},
   type = {Journal Article}
}

@article{1977Marcus,
   author = {Marcus, R. A. and Coltrin, Michael E.},
   title = {A new tunneling path for reactions such as H+H2→H2+H},
   journal = {The Journal of Chemical Physics},
   volume = {67},
   number = {6},
   pages = {2609-2613},
   ISSN = {0021-9606},
   DOI = {10.1063/1.435172},
   url = {https://doi.org/10.1063/1.435172},
   year = {1977},
   type = {Journal Article}
}

@article{2019Meisner,
   author = {Meisner, J. and Kamp, I. and Thi, W.-F. and K{\"a}stner, J.},
   title = {The role of atom tunneling in gas-phase reactions in planet-forming disks},
   journal = {\aap},
   volume = {627},
   pages = {A45},
   url = {https://doi.org/10.1051/0004-6361/201834974},
   year = {2019},
   type = {Journal Article}
}

@article{2017Meisner,
   author = {Meisner, Jan and Lamberts, Thanja and Kästner, Johannes},
   title = {Atom Tunneling in the Water Formation Reaction H2 + OH → H2O + H on an Ice Surface},
   journal = {ACS Earth and Space Chemistry},
   volume = {1},
   number = {7},
   pages = {399-410},
   DOI = {10.1021/acsearthspacechem.7b00052},
   url = {https://doi.org/10.1021/acsearthspacechem.7b00052},
   year = {2017},
   type = {Journal Article}
}

@article{2013Minissale,
   author = {Minissale, M. and Congiu, E. and Baouche, S. and Chaabouni, H. and Moudens, A. and Dulieu, F. and Accolla, M. and Cazaux, S. and Manicó, G. and Pirronello, V.},
   title = {Quantum Tunneling of Oxygen Atoms on Very Cold Surfaces},
   journal = {Physical Review Letters},
   volume = {111},
   number = {5},
   pages = {053201},
   DOI = {10.1103/PhysRevLett.111.053201},
   url = {https://link.aps.org/doi/10.1103/PhysRevLett.111.053201},
   year = {2013},
   type = {Journal Article}
}

@article{2012Oba,
   author = {Oba, Y. and Watanabe, N. and Hama, T. and Kuwahata, K. and Hidaka, H. and Kouchi, A.},
   title = {Water Formation Through A Quantum Tunneling Surface Reaction, OH + H2, AT 10 K},
   journal = {The Astrophysical Journal},
   volume = {749},
   number = {1},
   pages = {67},
   ISSN = {0004-637X},
   DOI = {10.1088/0004-637X/749/1/67},
   url = {https://doi.org/10.1088/0004-637X/749/1/67},
   year = {2012},
   type = {Journal Article}
}

@article{1994Walker,
   author = {Walker, James S. and Gathright, J.},
   title = {Exploring one‐dimensional quantum mechanics with transfer matrices},
   journal = {American Journal of Physics},
   volume = {62},
   number = {5},
   pages = {408-422},
   ISSN = {0002-9505},
   DOI = {10.1119/1.17541},
   url = {https://doi.org/10.1119/1.17541},
   year = {1994},
   type = {Journal Article}
}

@article{2008Watanabe,
   author = {Watanabe, Naoki and Kouchi, Akira},
   title = {Ice surface reactions: A key to chemical evolution in space},
   journal = {Progress in Surface Science},
   volume = {83},
   number = {10},
   pages = {439-489},
   ISSN = {0079-6816},
   DOI = {https://doi.org/10.1016/j.progsurf.2008.10.001},
   url = {https://www.sciencedirect.com/science/article/pii/S0079681608000476},
   year = {2008},
   type = {Journal Article}
}

@article{2000HenkelmanCINEB,
   author = {Henkelman, G. and Uberuaga, B. P. and Jonsson, H.},
   title = {A climbing image nudged elastic band method for finding saddle points and minimum energy paths},
   journal = {Journal of Chemical Physics},
   volume = {113},
   number = {22},
   pages = {9901-9904},
   ISSN = {0021-9606},
   DOI = {10.1063/1.1329672},
   url = {<Go to ISI>://WOS:000165584900005},
   year = {2000},
   type = {Journal Article}
}

@article{2000Henkelmantangent,
   author = {Henkelman, Graeme and Jónsson, Hannes},
   title = {Improved tangent estimate in the nudged elastic band method for finding minimum energy paths and saddle points},
   journal = {Journal of Chemical Physics},
   volume = {113},
   number = {22},
   pages = {9978-9985},
   ISSN = {0021-9606},
   DOI = {10.1063/1.1323224},
   url = {https://doi.org/10.1063/1.1323224},
   year = {2000},
   type = {Journal Article}
}

@ARTICLE{2026ApJ..1001...43W,
       author = {{Wang}, Lile and {Long}, Feng and {Yang}, Haifeng and {Dong}, Ruobing and {Xu}, Shenzhen},
        title = "{Adsorption of Volatiles on Dust Grains in Protoplanetary Disks}",
      journal = {\apj},
         year = 2026,
        month = apr,
       volume = {1001},
       number = {1},
          eid = {43},
        pages = {43},
          doi = {10.3847/1538-4357/ae4de3},
archivePrefix = {arXiv},
       eprint = {2511.09481},
 primaryClass = {astro-ph.EP},
       adsurl = {https://ui.adsabs.harvard.edu/abs/2026ApJ..1001...43W}
}

@article{Bai_2009,
doi = {10.1088/0004-637X/701/1/737},
url = {https://doi.org/10.1088/0004-637X/701/1/737},
year = {2009},
month = {jul},
publisher = {The American Astronomical Society},
volume = {701},
number = {1},
pages = {737},
author = {Bai, Xue-Ning and Goodman, Jeremy},
title = {HEAT AND DUST IN ACTIVE LAYERS OF PROTOSTELLAR DISKS},
journal = {The Astrophysical Journal},
}

@ARTICLE{2020A&A...641A.149R,
       author = {{Rogantini}, D. and {Costantini}, E. and {Zeegers}, S.~T. and {Mehdipour}, M. and {Psaradaki}, I. and {Raassen}, A.~J.~J. and {de Vries}, C.~P. and {Waters}, L.~B.~F.~M.},
        title = "{Magnesium and silicon in interstellar dust: X-ray overview}",
      journal = {\aap},
         year = 2020,
        month = sep,
       volume = {641},
          eid = {A149},
        pages = {A149},
          doi = {10.1051/0004-6361/201936805},
archivePrefix = {arXiv},
       eprint = {2007.03329},
 primaryClass = {astro-ph.HE},
       adsurl = {https://ui.adsabs.harvard.edu/abs/2020A&A...641A.149R}
}

@ARTICLE{2004ApJ...609..826K,
       author = {{Kemper}, F. and {Vriend}, W.~J. and {Tielens}, A.~G.~G.~M.},
        title = "{The Absence of Crystalline Silicates in the Diffuse Interstellar Medium}",
      journal = {\apj},
         year = 2004,
        month = jul,
       volume = {609},
       number = {2},
        pages = {826-837},
          doi = {10.1086/421339},
archivePrefix = {arXiv},
       eprint = {astro-ph/0403609},
 primaryClass = {astro-ph},
       adsurl = {https://ui.adsabs.harvard.edu/abs/2004ApJ...609..826K}
}

@ARTICLE{2023A&A...670A..30P,
       author = {{Psaradaki}, I. and {Costantini}, E. and {Rogantini}, D. and {Mehdipour}, M. and {Corrales}, L. and {Zeegers}, S.~T. and {de Groot}, F. and {den Herder}, J.~W.~A. and {Mutschke}, H. and {Trasobares}, S. and {de Vries}, C.~P. and {Waters}, L.~B.~F.~M.},
        title = "{Oxygen and iron in interstellar dust: An X-ray investigation}",
      journal = {\aap},
         year = 2023,
        month = feb,
       volume = {670},
          eid = {A30},
        pages = {A30},
          doi = {10.1051/0004-6361/202244110},
archivePrefix = {arXiv},
       eprint = {2210.05778},
 primaryClass = {astro-ph.HE},
       adsurl = {https://ui.adsabs.harvard.edu/abs/2023A&A...670A..30P}
}

@article{2025Jubert,
   author = {Jubert, Léana and Martínez-Bachs, Berta and Pareras, Gerard and Rimola, Albert},
   title = {Energy partitioning in H2 formation on interstellar carbonaceous grains. Insights from ab initio molecular dynamics simulations},
   journal = {Physical Chemistry Chemical Physics},
   volume = {27},
   number = {29},
   pages = {15385-15397},
   ISSN = {1463-9076},
   DOI = {10.1039/D5CP01585E},
   url = {http://dx.doi.org/10.1039/D5CP01585E},
   year = {2025},
   type = {Journal Article}
}
\bibliographystyle{aasjournalv7}

%



\appendix


\section{Atomic slab model construction} \label{sec:model}

Carbonaceous dust grains exist in various forms---graphite,
amorphous carbon, and organic solids---and simulating all
types with NQEs at full physical consistency is
infeasible. For catalytic properties, however, these
substrates are dominated by the surface carbon layer
\citep{Draine_book}. We therefore choose graphene as a
carbonaceous proxy. Hydrogen formation is studied on a
($4\times 4$) graphene slab with periodic boundary
conditions (PBC), containing 32 carbon atoms with two
adsorbed \chem{H^*} representing the initial state of two-H
association, and a $13~\ang$ vacuum region to suppress
spurious slab--slab interactions (Figure
\ref{fig:structure}, panel~a). The elementary steps modeled
are hydrogen adsorption/desorption, \chem{H^*} hopping, and
two-H association from three distinct configurations (ortho,
meta, para; Figure \ref{fig:structure}).

Silicates exhibit diverse stoichiometry and are
predominantly amorphous. They were once thought to be
virtually crystalline-free based on infrared upper limits
($<0.5\%$ in the diffuse~ISM;
\citealt{2004ApJ...609..826K}), but X-ray fine-structure
absorption measurements have since established a crystalline
fraction of $\sim 10\%$ along many sight lines
\citep{2020A&A...641A.149R, 2023A&A...670A..30P}. Following
previous computational studies
\citep[e.g.][]{2026ApJ..1001...43W}, we choose iron-poor
enstatite (Pnma-\chem{MgSiO_3}) as a silicate proxy, since
surface reactivity is governed by local coordination rather
than long-range order. We build a ($2\times 2$)
\chem{MgSiO_3} (001) slab with three layers: a nonpolar Mg-O
termination on top and a nonpolar O-Si-O termination on the
bottom, with the middle layer fixed (Figure
\ref{fig:structure}, panel~c). A $15~\ang$ vacuum region and
PBC are applied. Among four candidate sites (O, Mg on the
Mg-O termination; O, Si on the O-Si-O termination), the O
site on the Mg-O termination is most stable, with energy
$\sim 0.5$~eV lower than the other sites (Table
\ref{tab:site}; Appendix~\ref{sec:dft}). These O sites form
quasi-one-dimensional chains: intra-chain spacing
$\sim 2.91~\ang$ and inter-chain spacing $\sim 4.31~\ang$,
so we model motion only within a single chain. The
elementary steps considered are adsorption/desorption,
\chem{H^*} hopping, and two-H association (Figure
\ref{fig:structure}, panel~d).

  Our multi-scale simulation paradigm (combining
  DFT, free energy calculations, and KMC) requires periodic,
  low-defect crystalline surfaces as input, reflecting a
  practical constraint shared by these methods. Simulating
  amorphous substrates (silicates or carbon) with equivalent
  quantum accuracy is computationally prohibitive at
  present; converged free energy barriers for H adsorption,
  diffusion, and association over a statistically meaningful
  ensemble of local environments also remain out of
  reach. Because barriers for these steps vary across local
  configurations, each distinct environment would require
  its own parametrization, leading to a combinatorial
  explosion of possible sites and pathways.  We therefore
  present these systems as proof-of-concept models that
  demonstrate the method's ability to capture NQEs across
  two chemically distinct grain classes, and as meaningful
  (though incomplete) samplings and proxies for the
  amorphous surfaces likely relevant in astrophysical
  environments.

\section{Density functional theory (DFT) calculation setups
  and results} \label{sec:dft}

In this work, DFT \citep{Kohn1965DFT} calculations are used
to label surface configurations and prepare training
datasets for machine learning force fields (MLFFs). We
employ the first-principles calculation package: Vienna Ab
initio Simulation Package (VASP) \citep{Kresse1996CMSvasp,
  Kresse1996PRBvasp} with the generalized gradient
approximation (GGA) for the exchange-correlation functional
in the form of the Perdew-Burke-Ernzerhof (PBE) version
\citep{Perdew1996PRLpbe} and the projector augmented-wave
(PAW) method for pseudopotentials
\citep{Bloechl1994PRBpaw}. Convergence tests for the
graphene and \chem{MgSiO_3} slab models establish energy
convergence within 0.6 and 0.001~meV per atom, respectively,
when increasing the k-mesh from $3\times 3\times 1$ to
$4\times 4\times 1$, and within 2.2 and 2.0~meV per atom
when increasing the cutoff from 500 to 600~eV. Consequently,
a plane-wave cutoff of 500~eV and a $3\times 3\times 1$
$k$-mesh are adopted for all subsequent DFT calculations. We
employ the gaussian smearing method with a width of 0.1~eV
and 0.03~eV for graphene and \chem{MgSiO_3} respectively,
and add a $z$-direction dipole correction to eliminate the
interactions between two adjacent slab models under the
PBC. Since GGA functionals have difficulties in describing
long-range van der Waals forces \citep{Dobson2002AJC}, a
dispersion correction to the total energy is added according
to the DFT-D3 method \citep{Grimme2010JCPD3}.

  We acknowledge that PBE is not the most accurate
  functional for H adsorption on carbonaceous systems:
  \citet{2010_tunnel_H2formation} showed that hybrid
  meta-GGA functionals (e.g., MPWB1K) yield more reliable
  chemisorption barriers and binding energies on aromatic
  substrates. However, we have chosen PBE-D3 based on three
  practical considerations. First, our study focuses on the
  relative importance of NQEs and comparative trends across
  surface sites and temperatures, for which GGA functionals
  often provide qualitatively reliable barrier topologies.
  Second, PBE-D3 (along with related GGA-D3 functionals such
  as PBEsol-D3) has been widely adopted in computationally
  demanding surface astrochemistry, including on-the-fly
  instanton calculations of \chem{H_2} formation
  \citep{HanEtAl2022JPCL}, \textit{ab initio} molecular
  dynamics of energy partitioning \cite{2025Jubert}, and
  diffusion studies with tunneling corrections
  \citep{Tong2024JPCC}; this reflects the necessary trade-off
  between accuracy and computational cost when free-energy
  sampling and path-integral calculations are required.
  Third, the cost of hybrid meta-GGA functionals would be
  prohibitive for training MLFFs, free energy calculations,
  and PIMC simulations, each integral to our multi-scale
  workflow. We expect the qualitative conclusion that NQEs
  dramatically enhance \chem{H_2} formation at low
  temperatures to be robust with respect to the choice of
  functional, as tunneling probabilities depend primarily on
  barrier width and shape. Quantitative rate constants, however,
  would benefit from future benchmark studies with higher-level
  functionals.

The convergence criteria for self-consistent field
calculations and structure optimization are rigorously set
at $10^{-6}~\eV$ and $0.03~\eV~\ang^{-1}$. We test the
necessity of spin-polarized calculations for all elementary
steps at the graphene and \chem{MgSiO_3} surfaces. We
perform spin-polarized DFT calculations for labeling the
training dataset of hydrogen adsorption at the graphene
surface and all elementary steps at the \chem{MgSiO_3}
surface which exhibit energetic difference between
spin-polarized and non-spin-polarized setups.

We directly compare the stability of a \chem{H^*} atom at
four different sites at the \chem{MgSiO_3} surface (the Mg
site, the Si site and the O site at the Mg-O or O-Si-O
terminated surface) by calculating the relative energies of
these different adsorbed configurations
($E_{\mathrm{site-H}}$) since they share the same substrate
materials and the same adsorbate species. The results are
shown in Table \ref{tab:site}, which indicates that the most
favorable site for \chem{H^*} is the O site at the Mg-O
terminated surface.  We compute the potential energy barrier
for elementary steps using the climbing image nudged elastic
band (CI-NEB) method \citep{2000HenkelmanCINEB,
  2000Henkelmantangent} with five intermediate images by
VASP with the same settings described above. The results are
shown in Figure \ref{fig:CINEB-adsorp}.

\section{MLFFs training process and accuracy tests}
\label{sec:mlffs}

The MLFFs used in this work are constructed by the DP-GEN
\citep{ZhangEtAl2020DPGEN} workflow, which contains a series
of iterations to automatically explore and label the
configurational space. Each iteration consists of three
steps: (1) training the MLFFs based on the current dataset;
(2) exploration of the configurational space; (3) labeling
the candidate configurations by DFT calculations, which are
subsequently added to the dataset for the next training
loop. In the model training process, we use the DeePMD-kit
software \citep{WangEtAL2018DPMDKit} to train four different
models based on the same dataset with different random seeds
used for the parameters' initialization. The neural network
contains an embedding network with three layers consisting
of 25, 50 and 100 nodes and a fitting network with three
layers consisting of 240 nodes for each layer. The learning
rate exponentially decays from $10^{-3}$ to
$3.51 \times 10^{-8}$ when minimizing the loss function for
all the models. During the exploration step, one of the four
MLFFs is chosen in the software LAMMPS \citep{Lu2021CPCdpmd,
  Thompson2022CPClammps} to perform the enhanced
sampling. For the configurations in the exploration
trajectories, the maximal standard deviation (we call it the
``model deviation'') is calculated based on the atomic
forces predicted by the four MLFFs \citep{Zhang2019PRM}. We
set an upper and lower bound (noted as $t_{\mathrm{hi}}$ and
$t_{\mathrm{lo}}$ respectively) of the trust level and the
candidate configurations are selected if their model
deviations fell within the bounds. At the labeling step, the
candidate configurations are computed by first-principles
calculations and added to the dataset for the next training
loop. If the model deviations of more than 90\% of all the
structures in a $10^4$-step trajectory are smaller than
$t_{\mathrm{lo}}$, the DP-GEN loops are regarded as
converged. For graphene substrates, $t_{\mathrm{lo}}$ and
$t_{\mathrm{hi}}$ are set in the ranges of
$0.05-0.1~\eV~\ang^{-1}$, and $0.15-0.3~\eV~\ang^{-1}$,
respectively, for different elementary steps. For the
\chem{MgSiO_3} cases, $t_{\mathrm{lo}}$ and
$t_{\mathrm{hi}}$ varies in the ranges of
$0.1 - 0.2~\eV~\ang^{-1}$, and $0.3-0.4~\eV~\ang^{-1}$,
respectively.

\begin{figure*}[ht!]
\includegraphics[width=0.7\linewidth]{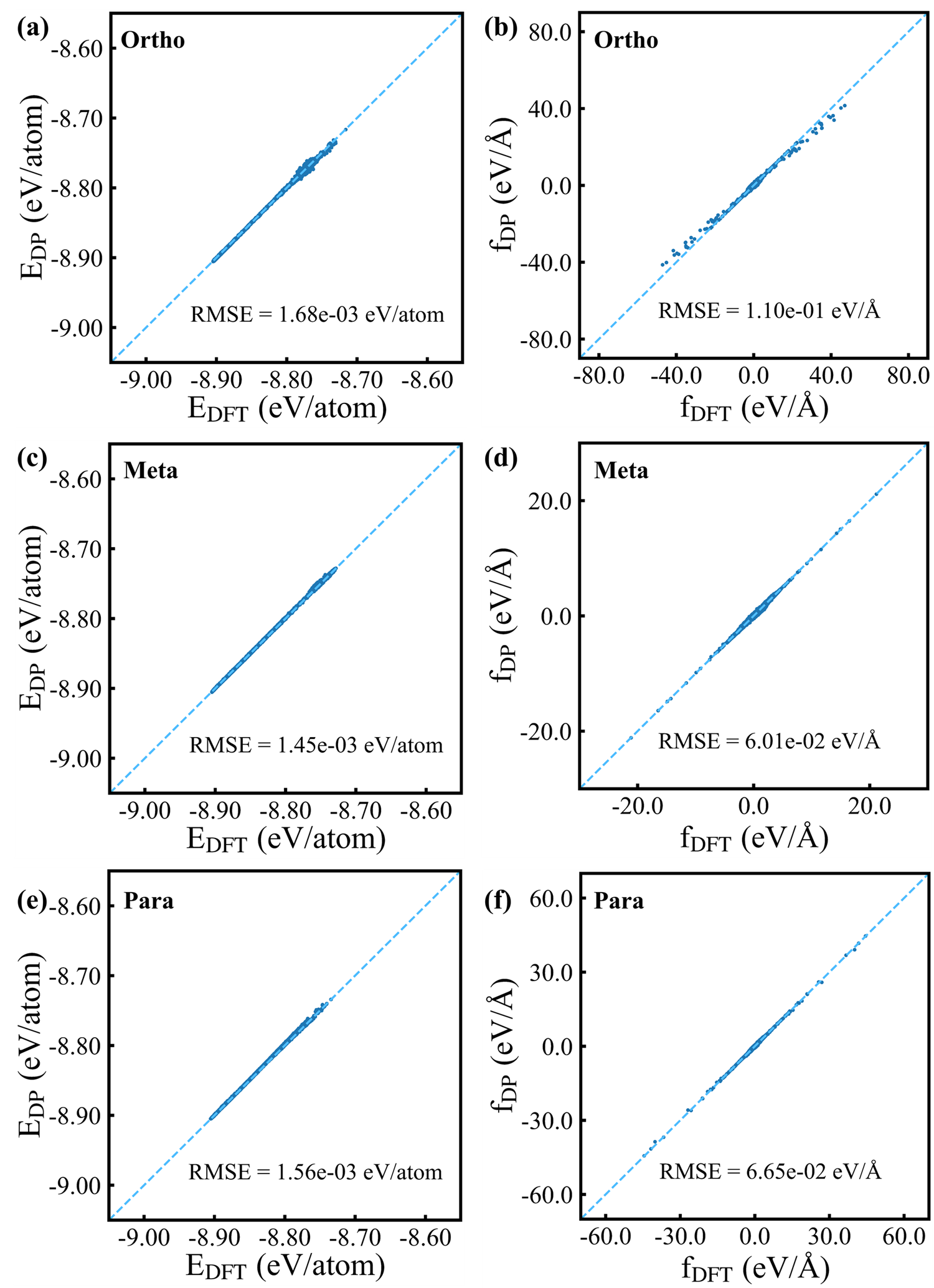}
\centering
\caption{Comparisons of energies and forces obtained by our
  MLFFs and the DFT calculations on the testing dataset of
  the two-H association elementary steps at the graphene
  surface. Panels (a) and (b) are the results of the ortho
  step with 2695 test data, panels (c) and (d) are the
  results of the meta step with 2400 test data, and panels
  (e) and (f) correspond to the para step with 3899 test
  data.  
  \label{fig:dp_her}}
\end{figure*}

\begin{figure*}[ht!]
\includegraphics[width=0.7\linewidth]{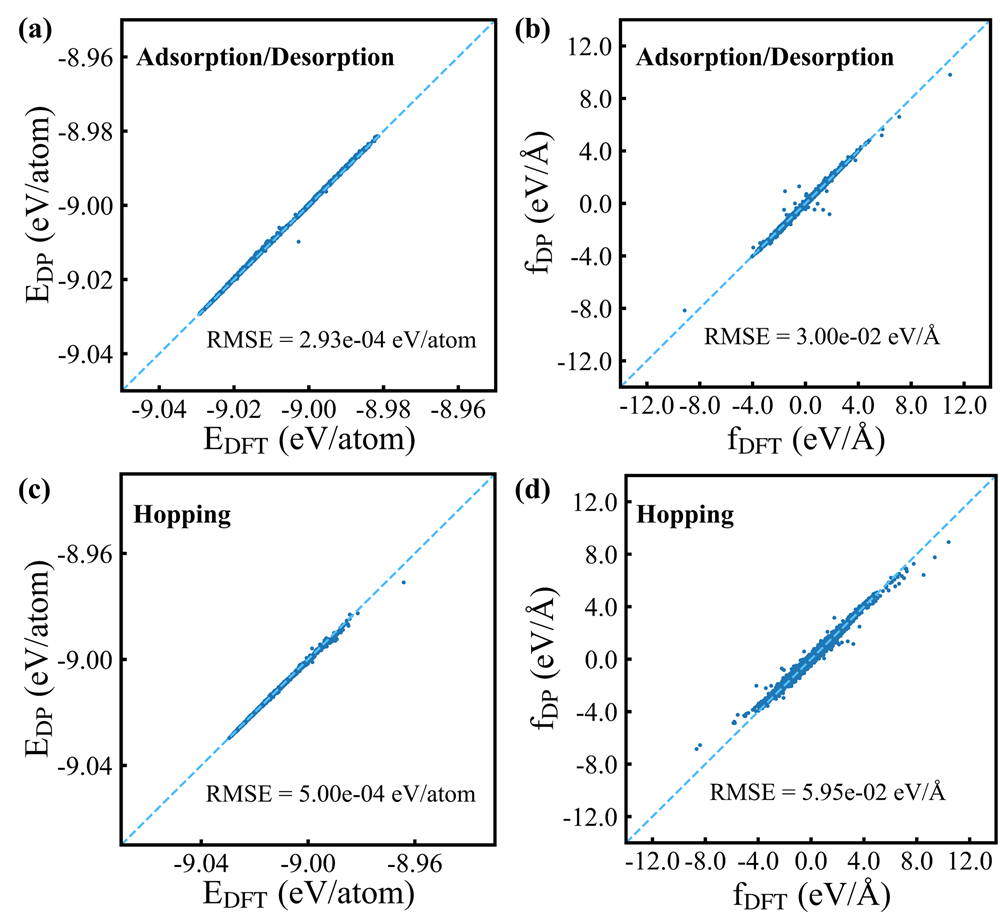}
\centering
\caption{Similar to Figure \ref{fig:dp_her}, but illustrates
  the results of (a-b) hydrogen adsorption and desorption with
  3300 test data and (c-d) hopping with 1800 test data at
  the graphene surface. 
  \label{fig:dp_hop}}
\end{figure*}

\begin{figure*}[ht!]
\includegraphics[width=0.7\linewidth]{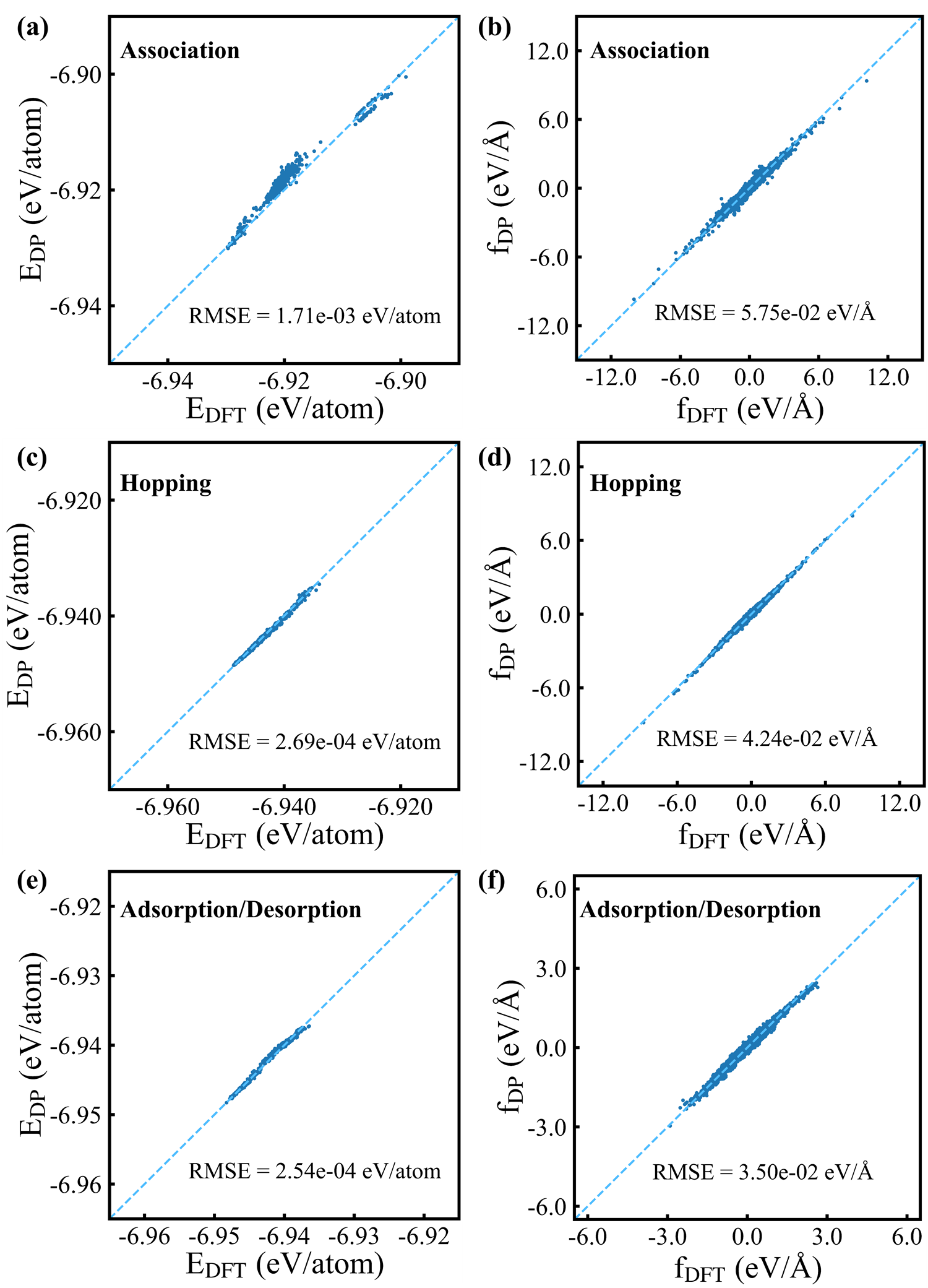}
\centering
\caption{Similar to Figure \ref{fig:dp_her}, but illustrates
  the results of (a-b) two-H association step with 450 test
  data, (c-d) hydrogen hopping step with 540 test data and
  (e-f) hydrogen adsorption and desorption step with 540 test
  data at the \chem{MgSi O_3} surface.
\label{fig:dp_mg}}
\end{figure*}

To test the accuracy of the MLFFs, we randomly select
structures from our quantum statistical sampling
calculations as the testing dataset. Energies and forces
inferred by the MLFFs matches well with the ones from the DFT
calculations on the testing dataset for all the elementary
steps, including two-H association, \chem{H^*} hopping and
hydrogen adsorption and desorption, which are shown in
Figure \ref{fig:dp_her} - \ref{fig:dp_mg}.

\section{Reaction coordinates for the graphene and
  \chem{MgSiO_3} models}
\label{sec:rc}

We define appropriate reaction coordinates (RCs) for each elementary step to drive reactions and obtain the free energy profiles. Two types of RCs are used: interatomic distances for association and adsorption/desorption steps, and differences of distances for hopping between neighboring sites. In the association step, we naturally define the distance between two hydrogen atoms as the RC, both for the graphene and the \chem{MgSiO_3} surfaces.
\begin{equation}
q_{\mathrm{HER}} = \left|\boldsymbol{r}_{\mathrm{H_{1}H_{2}}}\right|.
\end{equation}

As for the adsorption/desorption reaction, we employ the RC as the distance between the H atom and the adsorption site. We use ``adsite'' to generally label the adsorption site in different slab models, which means the C atom in the graphene case and the O atom in the \chem{MgSiO_3} case. Since adsorption and desorption are the forward and reverse of the same elementary step, the same RC applies to both.
\begin{equation}
q_{\mathrm{ad/de}} = \left|\boldsymbol{r}_{\mathrm{H}-\mathrm{adsite}}\right|.
\end{equation}

In the hopping process, we use the second type of RCs. We only consider the hopping between two neighboring sites. The RC is defined as the signed difference between the distances to the two sites, which is positive when the atom is closer to site 1, zero at the symmetric transition state, and negative when closer to site 2. We employ the difference between $\left|\boldsymbol{r}_{{\mathrm{H-adsite}}_\mathrm{1}}\right|$ and $\left|\boldsymbol{r}_{{\mathrm{H-adsite}}_\mathrm{2}}\right|$ as the RC $q_{\mathrm{hop}}$, in which $\left|\boldsymbol{r}_{{\mathrm{H-adsite}}_\mathrm{1}}\right|$ is the distance between the adsorption site 1 and the hopping H atom, and $\left|\boldsymbol{r}_{{\mathrm{H-adsite}}_\mathrm{2}}\right|$ is the distance between the adsorption site 2 and the hopping H atom
\begin{equation}
q_{\mathrm{hop}} =\left|\boldsymbol{r}_{{\mathrm{H-adsite}}_\mathrm{1}}\right| - \left|\boldsymbol{r}_{{\mathrm{H-adsite}}_\mathrm{2}}\right|.
\end{equation}

When nuclear quantum effects are included via the path integral method \citep{Feynman1948PI}, each H atom is represented as a ring polymer of $P$ beads. The reaction coordinate is then defined using the centroid $\overline{\boldsymbol{r}}_i \equiv P^{-1}\sum_{k=1}^{P} \boldsymbol{r}_i^{(k)}$ of each quantized atom, which represents its position in the classical limit. The same functional forms defined above are used with $\boldsymbol{r}_i$ replaced by $\overline{\boldsymbol{r}}_i$, ensuring that the RC drives the centroid while the internal bead fluctuations are sampled separately via the staging algorithm. The RCs are illustrated in Figure \ref{fig:free_energy} and the resulting free energy profiles are discussed in Appendix \ref{sec:mg_free}. 

\section{Sampling method and free energy calculations} \label{sec:free_energy}

In this section, we introduce the methods used in this work to calculate free energy profiles of elementary steps considering NQEs.

\subsection{Thermodynamic integration (TI) by the constrained hybrid Monte Carlo (CHMC) method}

In the TI method \citep{Den2000JCP, Sprik1998JCP}, considering a system of $N$ particles with coordinates $\left\{\mathbf{R}_i\right\}_{i=1}^{N}$ (the index $i$ indicates the $i^{\mathrm{th}}$ particle), the canonical ensemble partition function $Z(N,\beta)$ can be expressed as:
\begin{equation}
Z\left(N,\beta\right)=C\int{d\mathbf{R}_1\cdots{d\mathbf{R}}_N\exp{\left[-\beta U\left(\left\{\mathbf{R}_i\right\}\right)\right]}},
\end{equation}
where $\beta=\frac{1}{k_\mathrm{B}T}$ is the inverse temperature, $k_\mathrm{B}$ is the Boltzmann constant, $C$ is the prefactor produced by the integral of momenta degrees of freedom and $U\left(\left\{\mathbf{R}_i\right\}\right)$ is the potential energy which is a function of the $N$ particles' coordinates. We define the RC as $q\left(\left\{\mathbf{R}_i\right\}\right)$, then the free energy gradient (we call it ``mean force'') at $q=s$ is calculated as:
\begin{equation}\label{eq:e5}
\left.\frac{\d F}{\d q}\right|_{q=s}=-\frac{1}{\beta P\left(s\right)}\frac{\d P\left(s\right)}{\d s},
\end{equation}
where $F$ is the free energy at $q\left(\left\{\mathbf{R}_i\right\}\right) = s$, and $P\left(s\right) = \left<\mathrm{\delta}\left(q\left(\left\{\mathbf{R}_i\right\}\right) - s\right)\right>_{\mathrm{canonical}}$  represents the system's probability density at $q = s$. Here $\left<\dots\right>_{\mathrm{canonical}}$ means the canonical ensemble average. Then the free energy difference between two given RCs $s_1$ and $s_2$ is obtained by integral:
\begin{equation}
  F\left(s_2\right)-F\left(s_1\right)=\int_{s_1}^{s_2}{\frac{\d F}{\d q}\d q}=\int_{s_1}^{s_2}\left<\left(\frac{\d F}{\d q}\right)_{\mathrm{estm}}\right>_{\mathrm{s}}^{\mathrm{cond}}\d q,
\end{equation}
where $\left<\dots\right>_{\mathrm{s}}^{\mathrm{cond}}$ stands for the conditional ensemble average at $q = s$. The CHMC \citep{Jin2023JCTC} method is used for mean force computation, which employs a coordinate transformation to separate degrees of freedom by decoupling the RC, $q\left(\left\{\mathbf{R}_i\right\}\right)$, thus keeping it fixed during sampling
\begin{equation}
\left\{\mathbf{R}_i\right\}_{i=1}^N{\longrightarrow}\boldsymbol{q}=\left(q\left(\left\{\mathbf{R}_i\right\}\right),\ \boldsymbol{q}_{\mathrm{trans}},\boldsymbol{q}_{\mathrm{primit}}\right),
\end{equation}
where $\boldsymbol{q}$ represents the system's coordinates after the transformation, $\boldsymbol{q}_{\mathrm{trans}}$ is the transformed coordinates related to the RC and $\boldsymbol{q}_{\mathrm{primit}}$ is the primitive coordinates unchanged in this transformation. The general form of $\left(\frac{\d F}{\d q}\right)_{\mathrm{estm}}$ can be derived from \ref{eq:e5} with the coordinate transformation \citep{Sun2025NC}:
\begin{equation}
\left(\frac{\d F}{\d q}\right)_{\mathrm{estm}}=\frac{\partial\widetilde{U}}{\partial q}-\kb T\frac{\partial}{\partial q}\ln\varGamma\left(\boldsymbol{q}\right),
\end{equation}
where $\widetilde{U}$ is the form of the potential energy after the coordinate transformation and $\varGamma(q)$ is the associated Jacobian. We sampled $\boldsymbol{q}_{\mathrm{primit}}$ and $\boldsymbol{q}_{\mathrm{trans}}$ using the hybrid Monte Carlo (HMC) \citep{Mehlig1992PRB:hmc} and conventional Metropolis scheme \citep{Metropolis1953JCP}, respectively, with all other degrees of freedom fixed in each case.

\subsection{The constrained path integral HMC method
  (CPIHMC)}

We integrate the path integral algorithm
\citep{Feynman1948PI} into the CHMC method for free energy
calculations involving NQEs, where a quantum particle is
mapped onto a ring-polymer model of beads coupled with
harmonic oscillators. The number of beads is labeled as
$P$ and the coordinates of the $k^{\rm th}$ bead of a
$N$-particle system are
$\left\{\mathbf{R}_i^{(k)}\right\}_{i=1}^N$. We denote the
quantum canonical partition function as
$Z_{\mathrm{qtm}}\left(N,\beta\right)$, which is similar
to $Z\left(N,\beta\right)$ of the classical situation
, except that the potential energy
$U\left(\left\{\mathbf{R}_i\right\}\right)$ is replaced by
$U_{\mathrm{eff}}\left(\left\{\mathbf{R}_i^{(k)}\right\}\right)$.
\begin{equation}
\begin{split}
Z_{\mathrm{qtm}}\left(N,\beta\right)
  ={}&\lim_{P\to\infty}C_P
       \int\prod_{i=1}^N\d\mathbf{R}_i^{(1)}\cdots
       \d\mathbf{R}_i^{(P)} \\
     &\qquad \times \exp\left\{ -\beta U_{\mathrm{eff}}\left(\left\{\mathbf{R}_i^{(k)}\right\}\right) \right\},
\end{split}
\end{equation}

\begin{equation}
\begin{split}
U_{\mathrm{eff}} \left(\left\{\mathbf{R}_i^{(k)}\right\}\right)
  ={}&\sum_{k=1}^P\sum_{i=1}^N
        \biggl[\frac{1}{2}m_i\omega_P^2
        \left(\mathbf{R}_i^{(k+1)}-\mathbf{R}_i^{(k)}\right)^2 \\
     &\qquad +\frac{1}{P}
        U\left(\left\{\mathbf{R}_i^{(k)}\right\}\right)
        \biggr]\Big|_{R_i^{(1)}=R_i^{(P+1)}},
\end{split}
\end{equation}
where $\omega_P =\sqrt P/\beta\hbar$ is the chain frequence
of the harmonic interaction between two adjacent beads, and
$C_P$ is the prefactor comes from the Gaussian integral of
momenta degrees of freedom in the PI method. For the quantum
system, we also need RC to drive the reactions, and the
centroid of the multiple beads is used to define RC. We
denote the centroid of multiple beads as $\mathcal{R}_{i}$
and RC $q=f(\mathcal{R}_i)$. We can decouple the
centroid through a coordinate transformation, then the
sampling strategy is similar with the classical case except
that the beads' relative positions are sampled through the
staging algorithm \citep{Tuckerman2023book} with
$\boldsymbol{q}_\mathrm{trans}$ and
$\boldsymbol{q}_\mathrm{primit}$ fixed.

We give a brief workflow of the CPIHMC method in Figure
\ref{fig:workflow}. The simulation starts with an initial
configuration $\left\{ R_i^{(k)}\right \}$ followed by a
trial move of two types of degrees of freedom (the centroid
of atomic coordinates, and internal degrees of freedom
within the quantized beads' configurations), which are
randomly chosen according to a random number $\eta$
satisfying a uniform distribution on [0, 1] at each step. In
the CHMC branch, we sample the centroids of particles to
explore the complex configurational space, while in the path
integral Monte Carlo branch, we sample the quantized beads'
configurations based on the staging algorithm to treat
NQEs. Subsequently, we evaluate estimators of concerned
physical quantities like the mean forces and potential
energies. This iterative sampling process continues 
until the required total MC steps are reached, and the ensemble
average of the physical quantities is calculated at the end
(see also \citealt{Jin2023JCTC} and \citealt{Sun2025NC} for
more details).

\subsection{Convergence of potential energies and mean
  forces in the free energy calculations}

\begin{figure*}[ht!]
\includegraphics[width=0.65\linewidth]{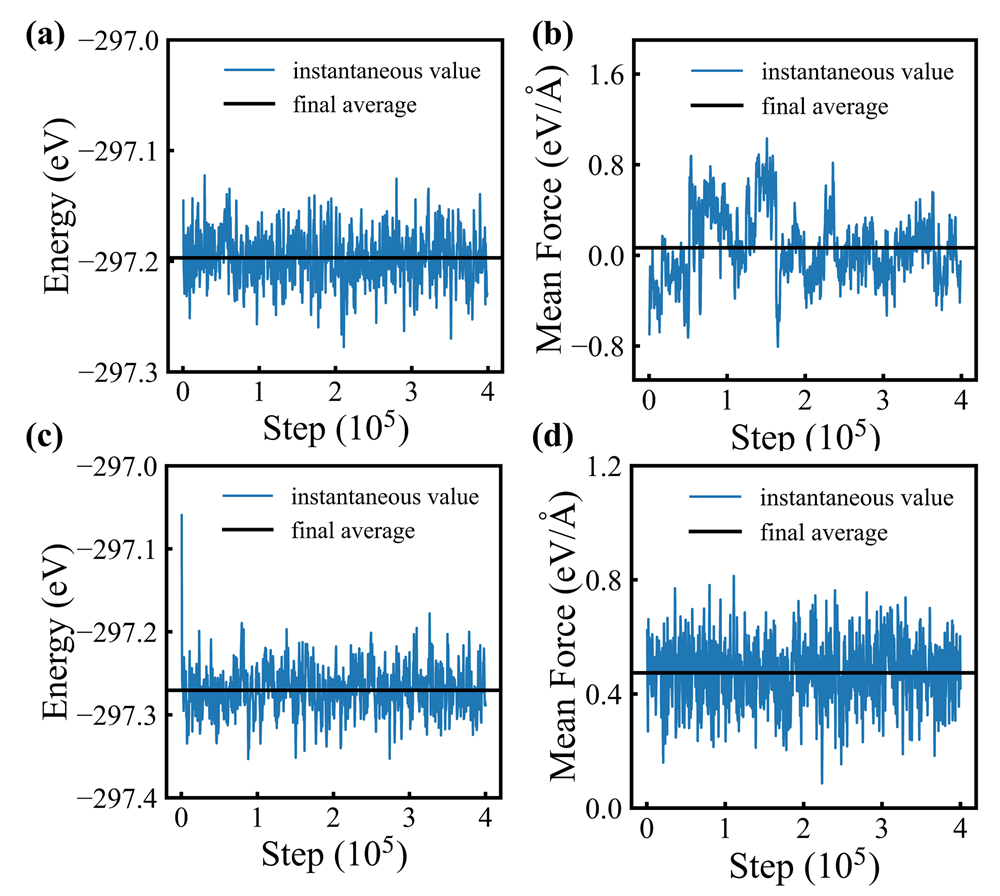}
\centering
\caption{Instantaneous fluctuation of the potential energies
  and mean forces along our CPIHMC sampling for (a-b) \chem{H^*}
  hopping step and (c-d) hydrogen adsorption step at the
  graphene surface. 
  \label{fig:conv_ad}}
\end{figure*}

\begin{figure*}[ht!]
\includegraphics[width=0.65\linewidth]{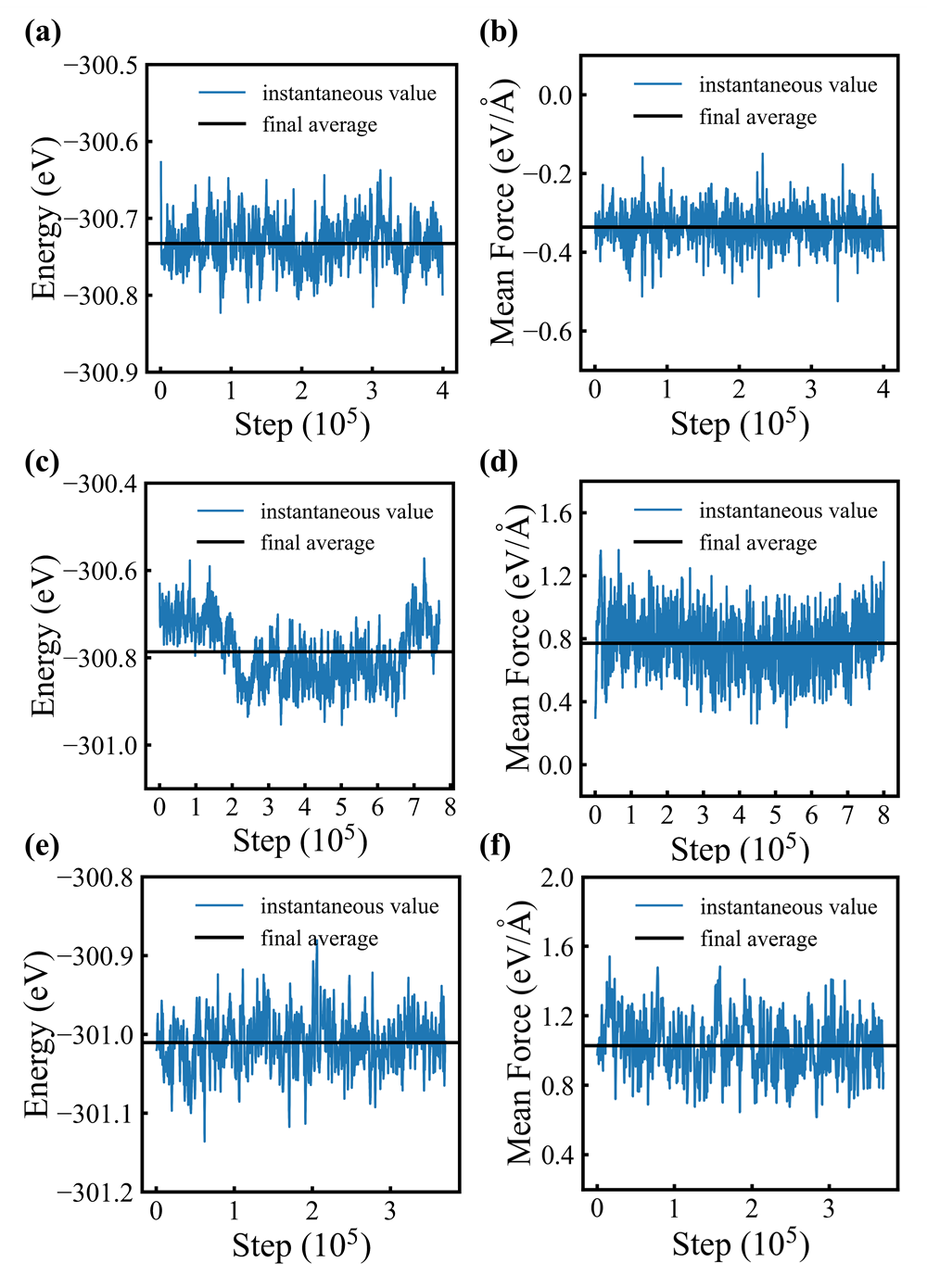}
\centering
\caption{Similar to Figure \ref{fig:conv_ad}, but
  illustrates the instantaneous fluctuation for two-H
  association steps at the graphene surface. (a-b) ortho
  step, (c-d) meta step, and (e-f) para step.
  \label{fig:conv_her}}
\end{figure*}

\begin{figure*}[ht!]
\includegraphics[width=0.65\linewidth]{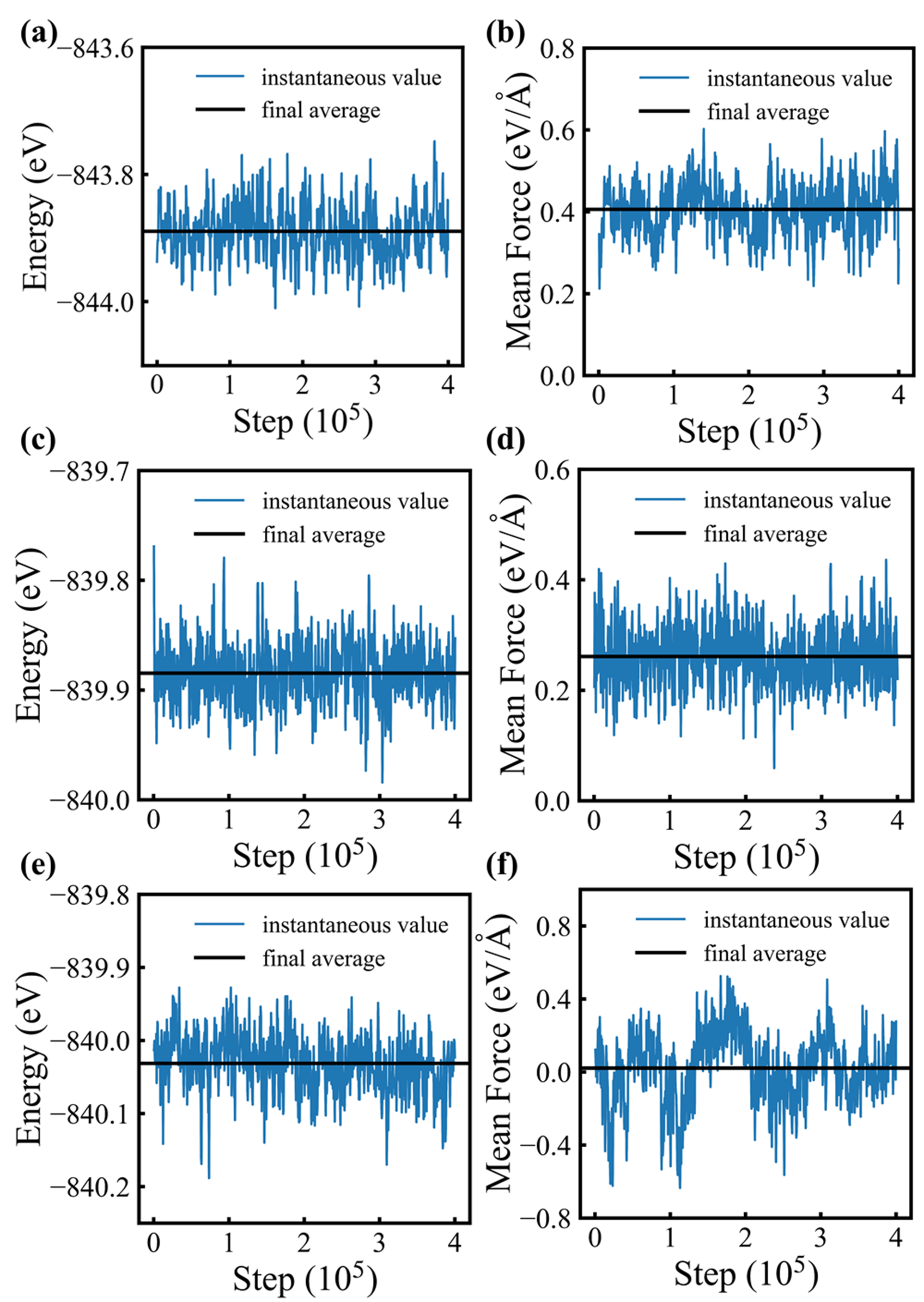}
\centering
\caption{Similar to Figure \ref{fig:conv_ad}, but
  illustrates the instantaneous fluctuation for elementary
  steps at the \chem{MgSiO_3} surface. (a-b) two-H
  association step, (c-d) hydrogen adsorption step and (e-f)
  \chem{H^*} hopping step. 
  \label{fig:conv_mg}}
\end{figure*}

We present the fluctuation of the mean force estimator and
the potential energy estimator near the transition state
(TS) in our sampling and the corresponding convergence
behavior in Figures \ref{fig:conv_ad} -
\ref{fig:conv_mg}. Since we find the CPIHMC sampling for the
quantum situation is more difficult to converge than the
CHMC case, we only present the CPIHMC results here.

\subsection{Mean force values along reaction coordinates}

\begin{figure*}[ht!]
\includegraphics[width=0.7\linewidth]{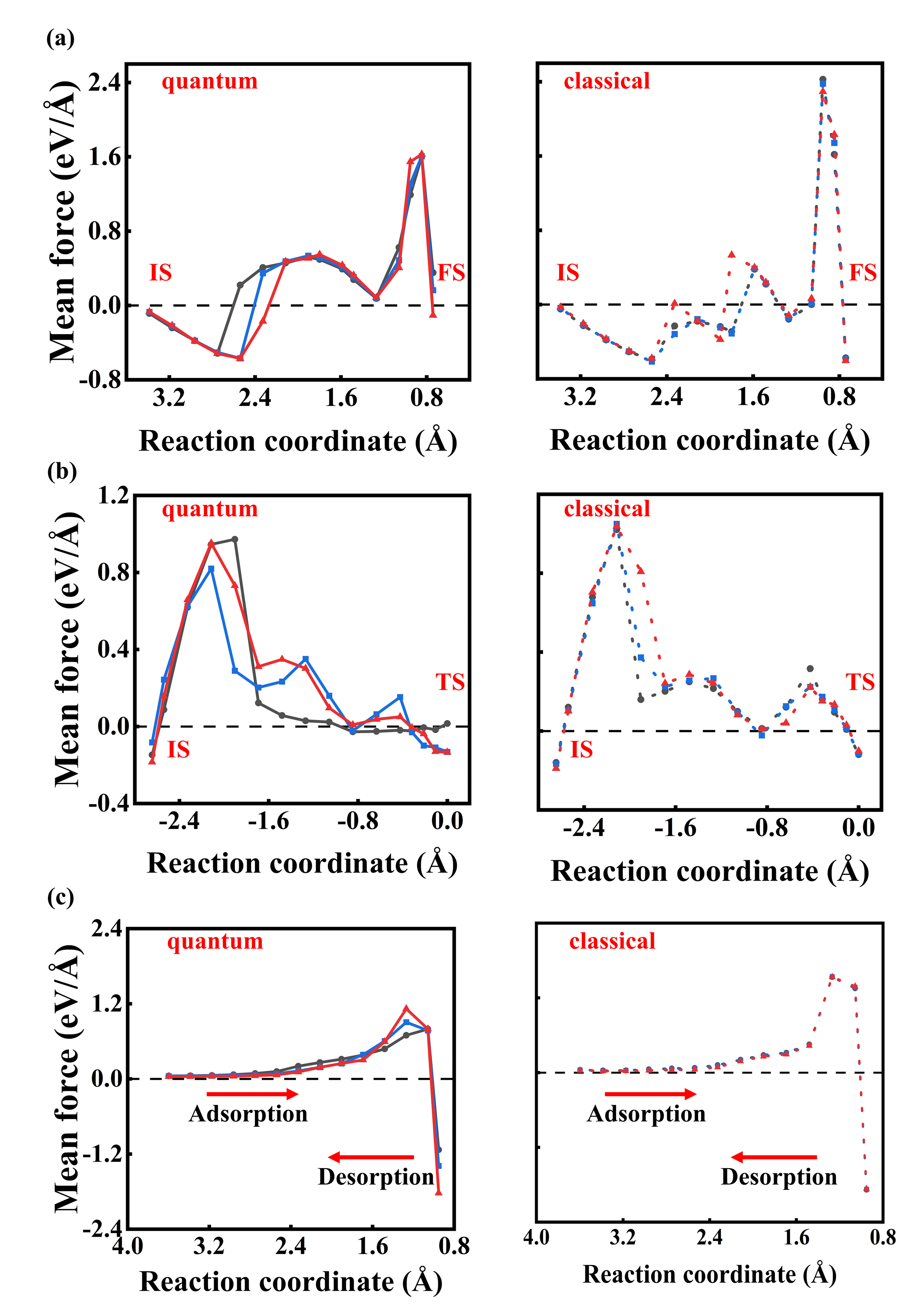}
\centering
\caption{Mean forces of the elementary steps along our
  defined RCs on the \chem{MgSiO_3} surface at temperatures
  of 50~K (black), 100~K (blue), and 200~K (red). The
  left panels stand for quantum mechanical results, and
  right panels the classical counterparts. Panels (a)
  for two-H association, (b) for hopping, and (c)
  for adsorption and desorption. 
\label{fig:mean_mg}}
\end{figure*}

\begin{figure*}[ht!]
\includegraphics[width=0.7\linewidth]{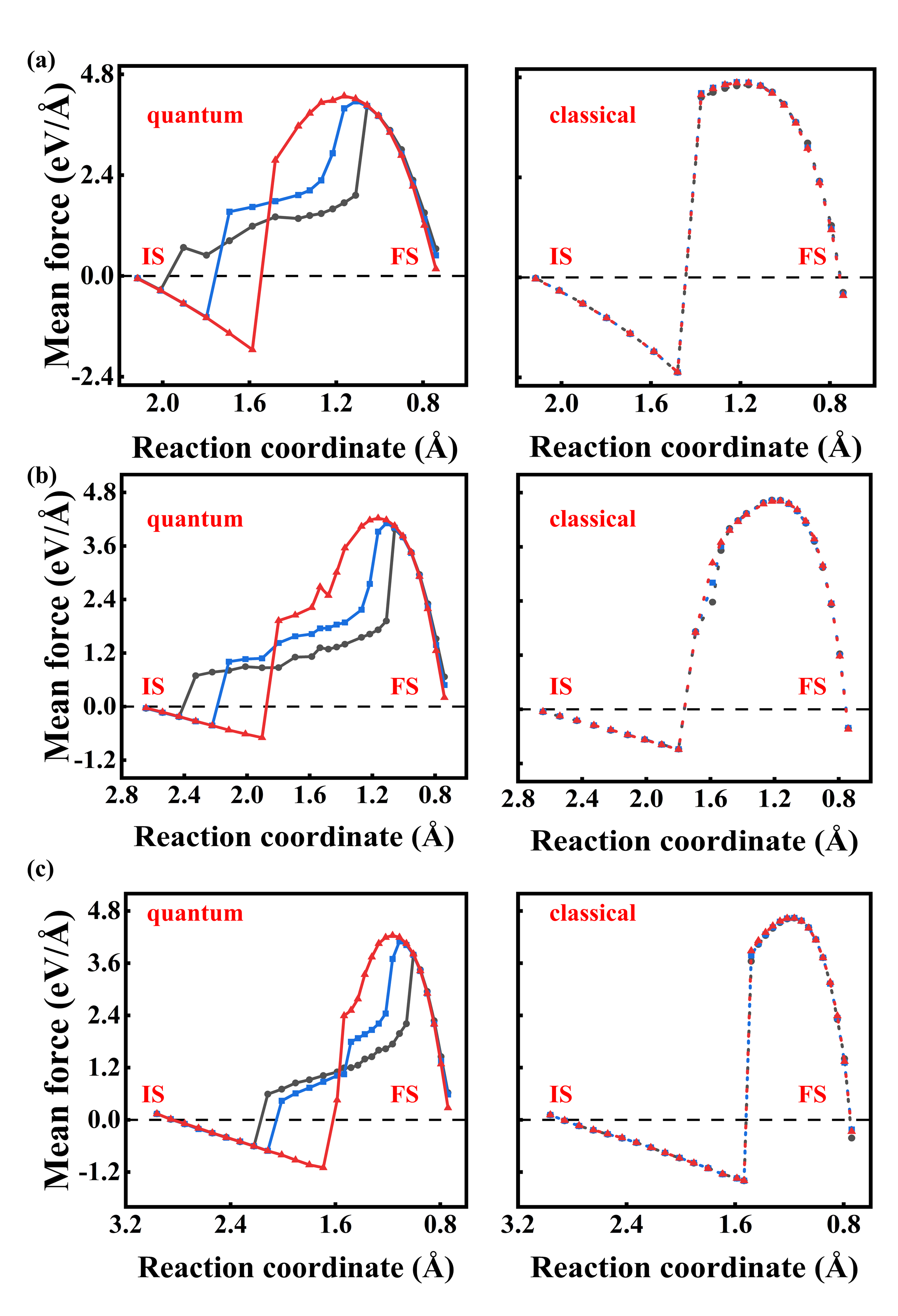}
\centering
\caption{Similar to Figure \ref{fig:mean_mg}, but
  illustrates the mean forces of (a) ortho step, (b) meta
  step and (c) para step on the graphene surface. 
\label{fig:mean_her}}
\end{figure*}

\begin{figure*}[ht!]
\includegraphics[width=0.7\linewidth]{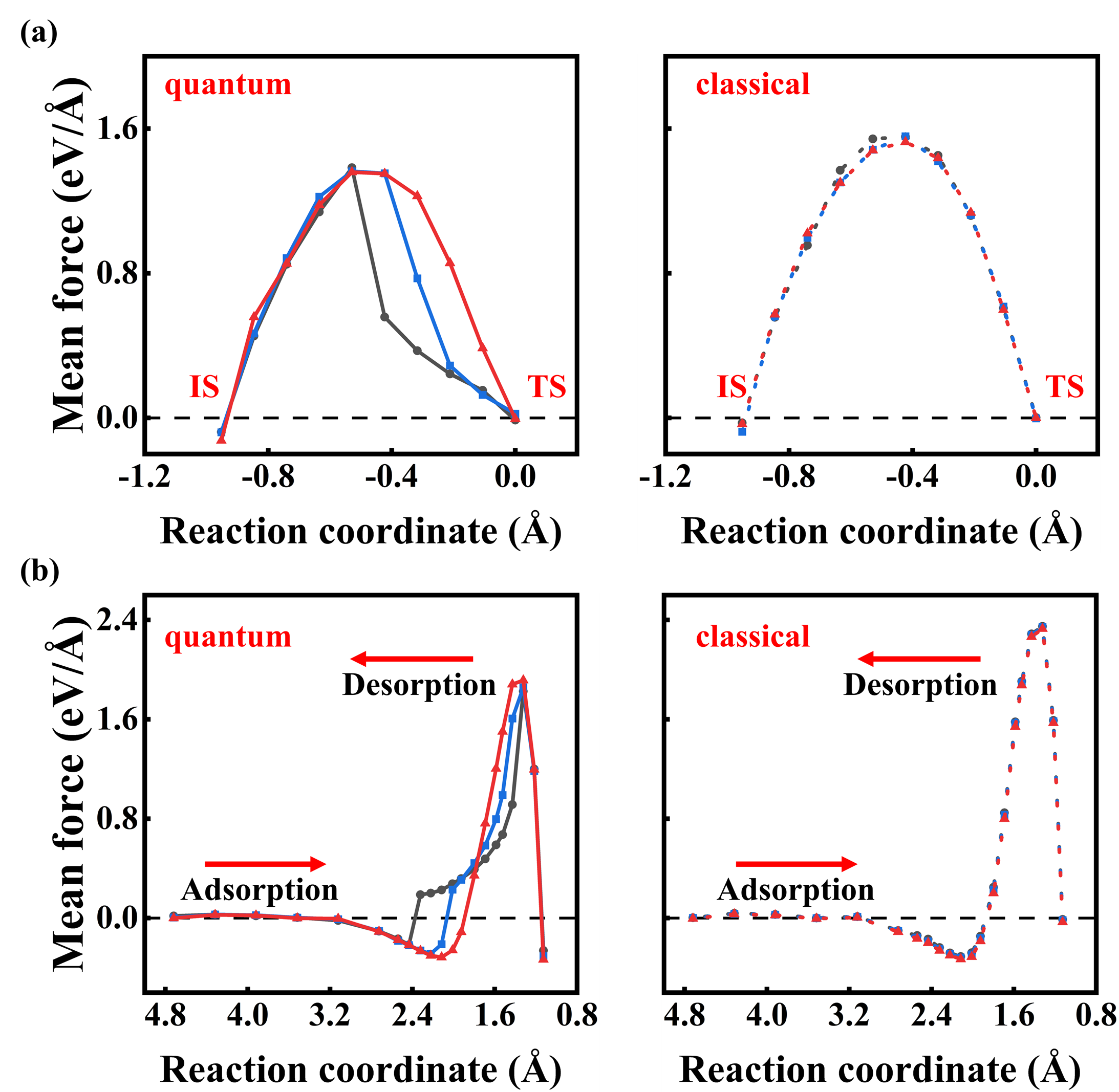}
\centering
\caption{Similar to Figure \ref{fig:mean_mg}, but
  illustrates the mean forces of (a) \chem{H^*} hopping and (b)
  hydrogen adsorption and desorption over the graphene surface. 
  \label{fig:mean_ad}}
\end{figure*}

In this section, we present the mean force values along our
defined RCs for both graphene and \chem{MgSiO_3} cases
sampled by the constrained (PI)HMC method
(Figure~\ref{fig:mean_mg} - \ref{fig:mean_ad}). For the
hopping step in both the graphene and \chem{MgSiO_3} cases,
we only show the results from the IS to the TS due to the
symmetric feature.

\subsection{Free energy results of the para step at the
  graphene surface and steps at the \chem{MgSiO_3} surface}
\label{sec:mg_free}

\begin{figure}[ht!]
\includegraphics[width=0.9\linewidth]{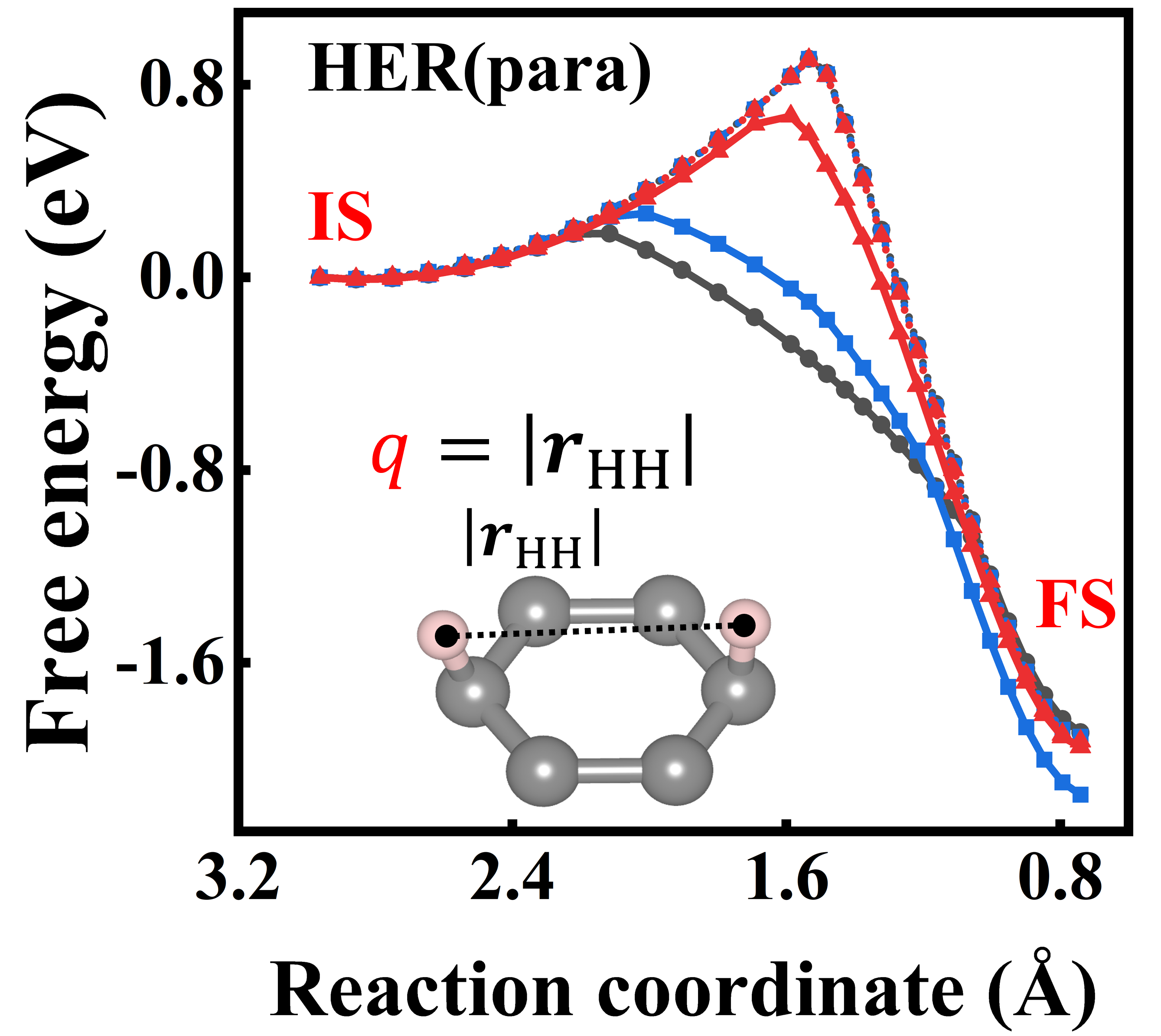}
\centering
\caption{Free energy profile of the two-H association (para)
  step on the graphene surface under the quantum (solid
  lines) and the classical (dotted lines) situations at
  temperatures of 50~K (black), 100~K (blue), and 200~K
  (red). 
\label{fig:free_para}}
\end{figure}

\begin{figure*}[ht!]
\includegraphics[width=0.6\linewidth]{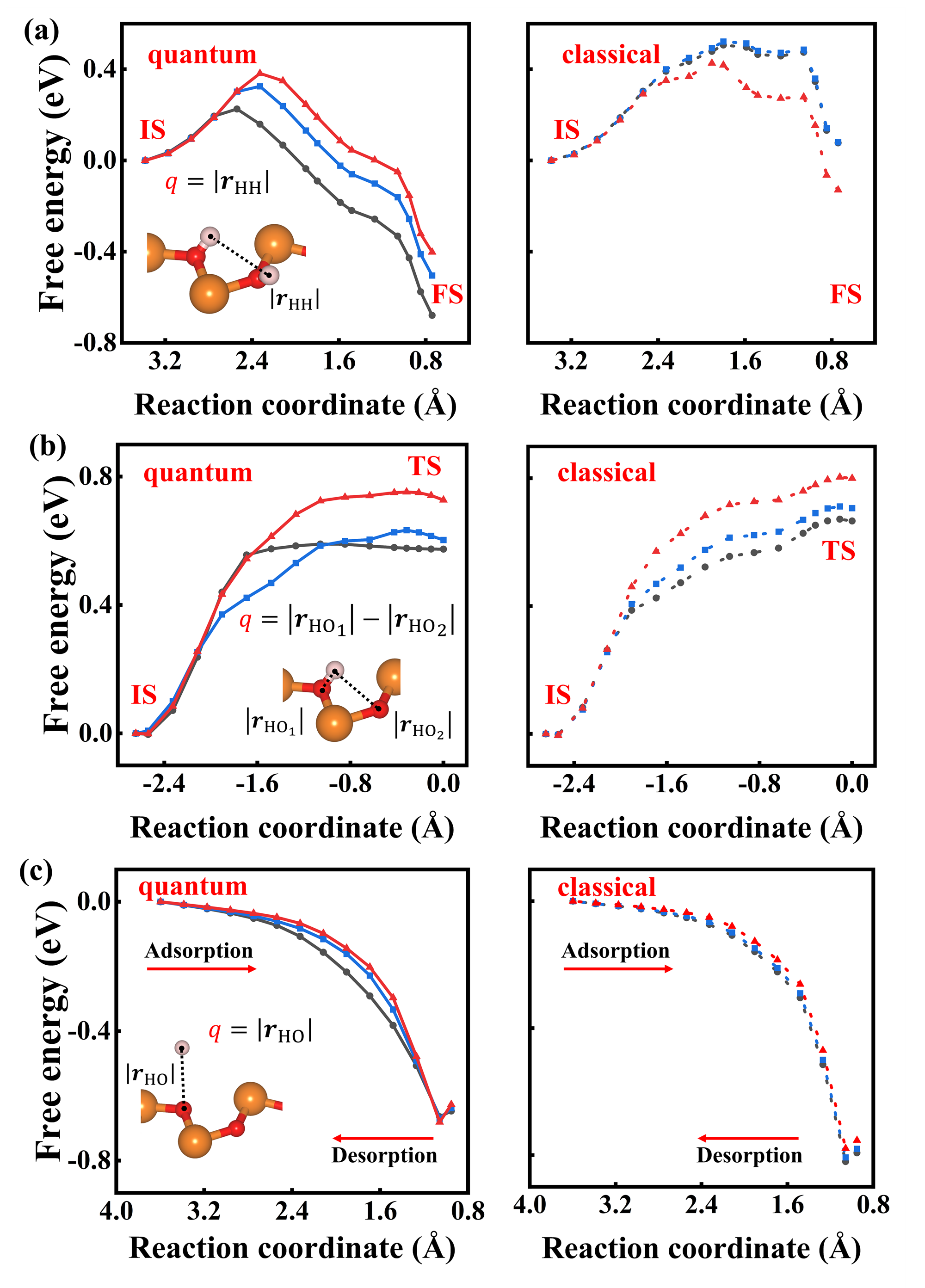}
\centering
\caption{Free energy profiles of the elementary steps on the
  \chem{MgSiO_3} surface at temperatures of 50~K (black),
  100~K (blue), and 200~K (red). The left and right panels
  present the quantum and classical results, respetively. Panels
   (a) for the two-H association step, (b) for the
   \chem{H^*} hopping step from the IS to the TS, and (c) for
  the hydrogen adsorption and desorption steps.  
  \label{fig:free_mg}}
\end{figure*}

This subsection shows free energy profiles of the para step
on the graphene surface (Figures~\ref{fig:free_para}) and
those on the \chem{MgSiO_3} surface
(Figures~\ref{fig:free_mg}).

\subsection{Graphs of the transition state with beads expansion}

\begin{figure*}[ht!]
\includegraphics[width=0.6\linewidth]{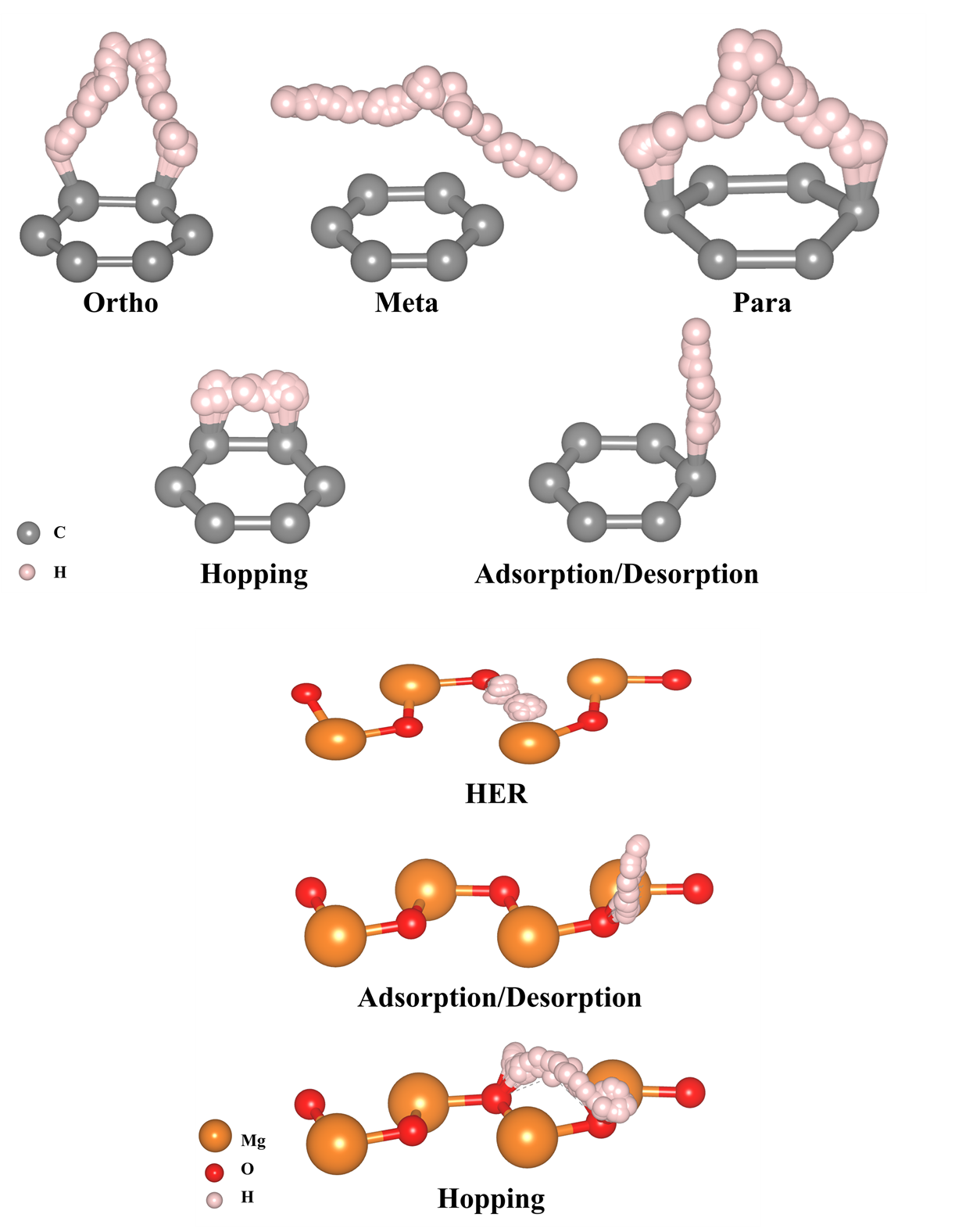}
\centering
\caption{The schematic structural plots of the TS with beads
  expansion in our sampling calcualtions for different steps
  on (a) the graphene surface and (b) the \chem{MgSiO_3}
  surface.  
\label{fig:beads}}
\end{figure*}

This section shows schematic snapshots of the TS with beads
expansion in all of the elementary steps (Figure~\ref{fig:beads}). We only consider
the expansion of the H atoms involved in the elementary
reaction steps.

\subsection{Miscellaneous setup details in the CPIHMC}

We includ some additional setups in our simulations for
multiple reasons. In CPIHMC simulations of the graphene
surface, the interactions between the near-final-state
``\chem{H_2} molecule'' and the surface are weak, so the
``\chem{H_2} molecule'' diffuses relatively freely in the
vacuum region. Sometimes, the molecule could go across the
upper bound of the slab model, reaching the other side of
the graphene surface due to the periodic boundary condition
(PBC). Thus we set a rigid wall in the vacuum region $\sim$
$6~\ang$ above the surface. If the near-final-state ``\chem{H_2}
molecule'' touches the wall, we reject this trial movement
in our sampling calculations. We fix the coordinate of one
carbon atom in the graphene layer to avoid the shifting
movement of the graphene along the z-direction.

In the \chem{MgSiO_3} case, we also adopt some
configuration restrictions during the sampling
calculations. First, our current MLFF occasionally
encounters an abnormal configuration during our CPIHMC
sampling simulations for the two-H association and
\chem{H^*} hopping elementary steps, in which a Mg or O atom
would leave the surface slab, leading to a failure of the
MLFF inference. We thus include a rigid wall for the Mg and
O atoms involved in hydrogen adsorption. The wall is
$\sim 0.2~\ang$ above the Mg atom in the z-direction,
stabilizing the geometry of the Mg-O terminated surface. In
addition, similar to the setup included in the graphene case
discussed above, we set a rigid wall for the
near-final-state ``\chem{H_2} molecule'' $\sim 7~\ang$ above
the surface slab.

\section{KMC simulations}
\label{sec:kmc}

\subsection{Workflow of the KMC calculations}

The KMC method is widely used to simulate the kinetic
properties based on an updating events list. Figure
\ref{fig:workflow} shows the workflow of KMC simulations,
which starts from an initial lattice structure. Global
events list is then constructed, which contains all the
possible elementary steps with rate constants. Determining
which event would occur and calculating the evolving time
step $\Delta t$ of the system are two key points in the KMC
method.

Suppose an events list at a certain KMC step contains $N$
possible elementary steps with their rate constants denoted
as $\left\{k_r\right\}_{r=1}^{N}$. Then the total rate
constant $k_\mathrm{tot}$ at the current KMC step is
constructed as:  
\begin{equation}
k_{\mathrm{tot}}=\sum_{r=1}^{N}k_r.
\end{equation}

The probability of the occurrence of the event r is
proportional to $k_r$, so that $P_r=k_r/k_\mathrm{tot}$. Two
random numbers $(\eta,\ \xi)$ according to a uniform
distribution within $[0, 1]$ are used to determine $j$ and
$\Delta t$
\begin{equation}
  \frac{1}{k_\mathrm{tot}}\sum_{r=1}^{j-1}k_r < \xi <
  \frac{1}{k_\mathrm{tot}}\sum_{r=1}^{j}k_r\, 
\end{equation}
\begin{equation}
\Delta t=-\frac{\ln{\left(\eta\right)}}{k_\mathrm{tot}},
\end{equation}
where $j$ is the index of the event being selected for
updating the system's configuration.

The selected event and the evolving time step are recorded
and the lattice structure is updated. Subsequently, a new
global events list is constructed and the above process is
repeated until reaching the required total KMC
steps. Finally, we can track the system's evolution and
statistically compute interested physical quantities.

\subsection{Reaction rate constants extrapolation}
\label{sec:extrapolation}

We do not directly compute free energy profiles at 20~K
because of the high computational cost associated with path
integral simulations, which require a larger number of beads
to achieve convergence at such low temperatures
\citep{Tuckerman2023book, Feynman1948PI}. Given that deep
tunneling dominates the elementary steps at 20~K, instead we
use an extrapolation formula Eq. \ref{Eq:exploration} to
obtain the reaction rate constants at 20~K using results at
50~K and 100~K \citep{1984Grabert, 1990Peter}:
 \begin{equation}
\label{Eq:exploration}
k(T) = k(0)\exp(aT^2).
\end{equation} 
We do not simulate the classical case at 20~K, as the
classical \chem{H_2} formation rate at 50~K is already
extremely low.

\begin{figure*}[ht!]
\includegraphics[width=0.8\linewidth]{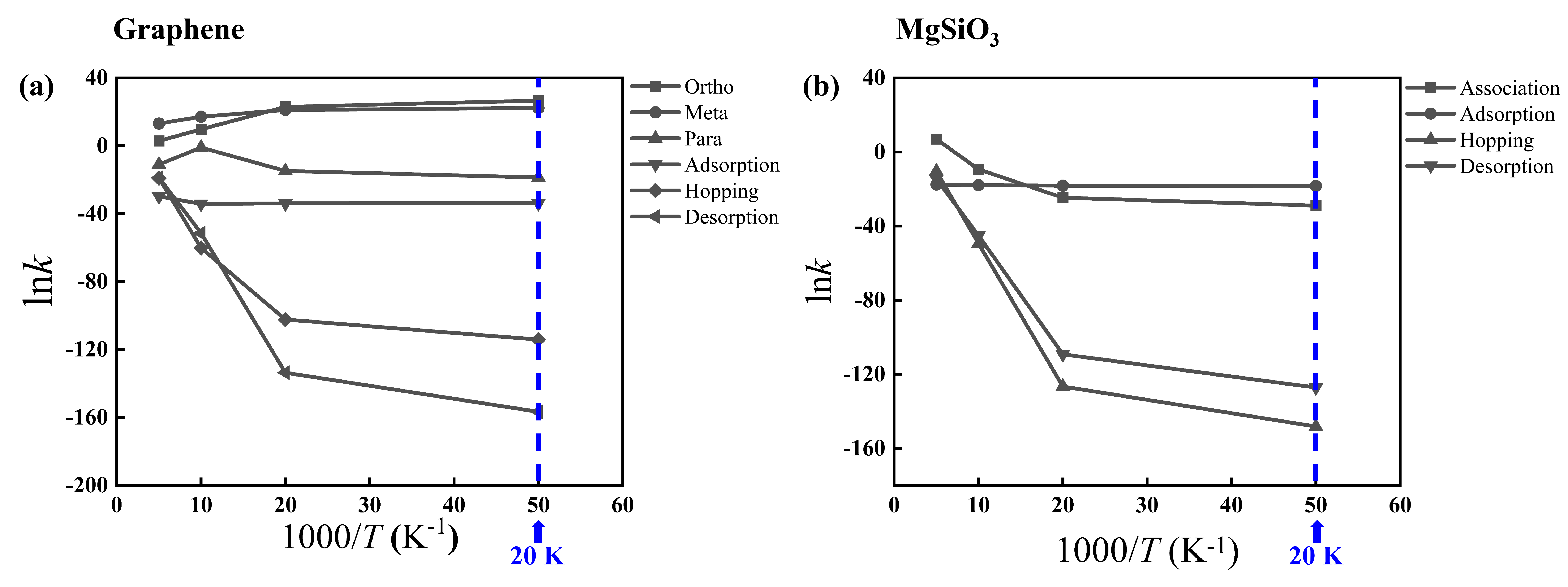}
\centering
\caption{Reaction rate constants of elementary steps as a
  function of temperature on (a) the graphene surface and
  (b) the \chem{MgSiO_3} surface. 
  \label{fig:rate_const}}
\end{figure*}

On the graphene surface, H adsorption is the rate-limiting
step at low temperatures, and its rate constant remains
nearly unchanged as temperature decreases from 50~K to
20~K. H hopping and desorption are frozen at low
temperatures, whereas the association steps (ortho and meta
configurations, illustrated in Figure \ref{fig:structure},
panel~b) are much faster than adsorption. Consequently, for
these non rate-limiting steps, whether we use the
extrapolation formula Eq. \ref{Eq:exploration} or simple
linear extrapolation has little impact on the overall
formation rate. This result is consistent with the
expectation that as temperature approaches zero, the
reaction rate constant should tend to a constant value. On
the \chem{MgSiO_3} surface, the rate-limiting step at low
temperatures is the two-H association step (illustrated in
Figure \ref{fig:structure}, panel~d). We also extrapolated
the corresponding rate constants to 20~K. The temperature
dependence of these rate constants on the graphene and
\chem{MgSiO_3} surface is shown in Figure
\ref{fig:rate_const}. The resulting \chem{H_2} formation
efficiency coefficients are presented in Figure
\ref{fig:kmc}. At 20~K, the \chem{H_2} formation rate on
graphene remains nearly unchanged from its value at 50~K,
while on \chem{MgSiO_3} it decreases slowly. The gap between
quantum and classical formation rates will be larger at 20~K
than at 50~K, indicating the importance of NQEs.

\subsection{Gas-dust temperature decoupling}
\label{sec:assumptions}

Two kinds of assumptions are taken when calculating the rate
constant of H adsorption: a thermalization assumption and an
adiabatic assumption. The thermalization assumption is
suitable for dense, cold environments
($T_{\mathrm{dust}} = T_{\mathrm{gas}}$), while the
adiabatic assumption is suitable for environments such as
PDRs ($T_{\mathrm{dust}} \gg T_{\mathrm{gas}}$). So, under
the adiabatic assumption, $T_{\mathrm{gas}}$ is used to
calculate the rate constant of H adsorption while
$T_{\mathrm{dust}}$ is used for other elementray
steps. Under this condition, $T_{\mathrm{gas}}$ is
sufficiently high that NQEs can be neglected and adsorption
is treated as H atoms overcoming a potential energy barrier
(rather than the free energy barrier used under the
thermalization assumption), with kinetic energies consistent
with $T_{\mathrm{gas}}$. The potential energy barrier is
calculated by CI-NEB as described in
Appendix~\ref{sec:dft}. The key difference between two
assumptions lies in the calculation of the adsorption rate
constant $k_{\mathrm{adsorp}}$. Under the adiabatic
assumption for H adsorption, H atoms do not thermalize with
the dust until after they are adsorbed; therefore,
$T_{\mathrm{gas}}$ rather than $T_{\mathrm{dust}}$ is used
in the $k_\mathrm{B}T$ factor of TST to compute
$k_{\mathrm{adsorp}}$.

\subsection{Coverage dependence}\label{sec:coverage}

\begin{figure*}[ht!]
\centering  
\includegraphics[width=0.7\linewidth]{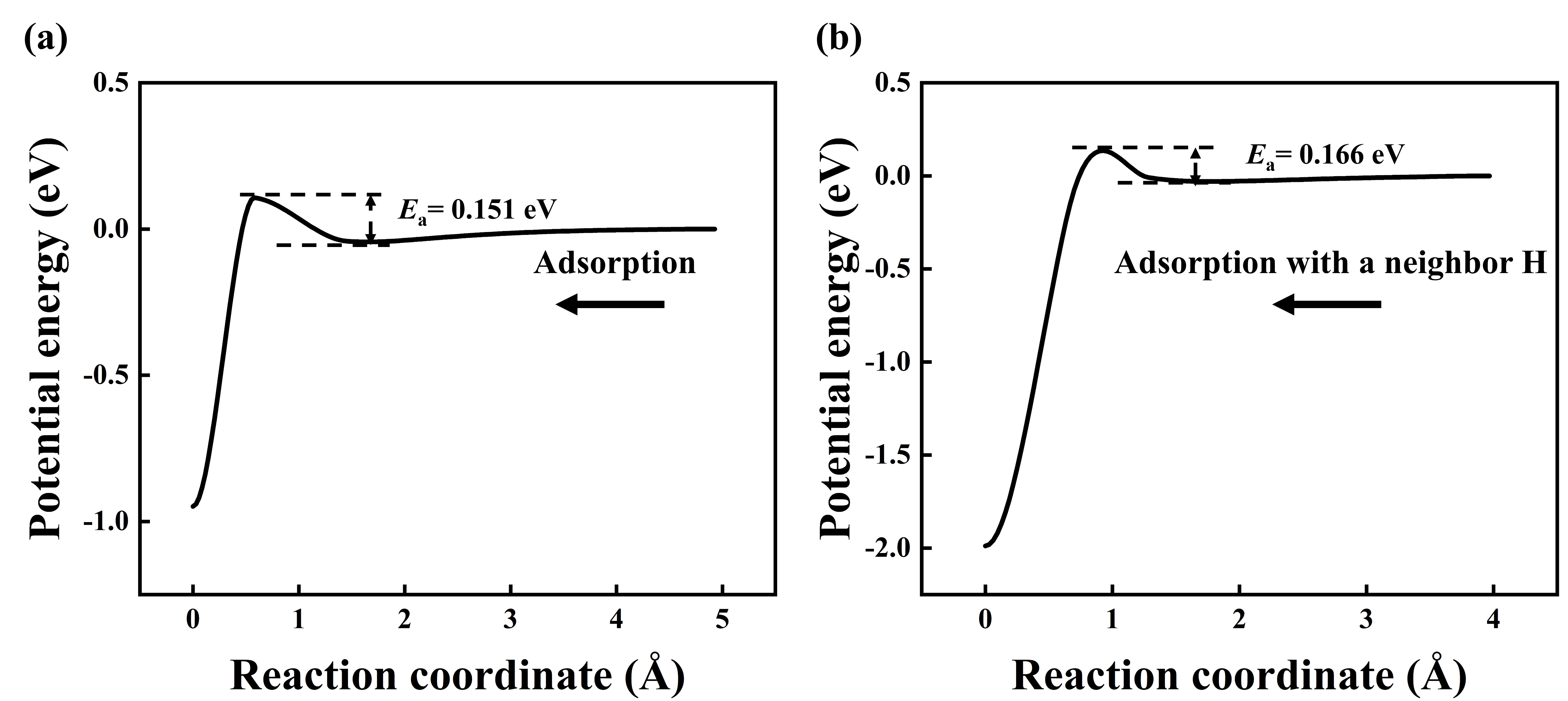}
\caption{Potential energy profile for H adsorption (a)
  without a neighboring adsorbed H atom, and (b) with a
  neighboring adsorbed H atom. 
  \label{fig:CINEB-adsorp}}
\end{figure*}

We perform additional DFT calculations to examine whether
the presence of adjacent H atoms modifies the adsorption
potential energy barrier for a new H atom on graphene. We
construct a $4\times 4$ supercell of graphene with one H
atom already chemisorbed on a carbon atom and then calculat
the barrier for a second H atom adsorbing on a neighboring
carbon atom. Our result shows that the adsorption barrier
for H on graphene changes only minimally (within
$\sim 0.015~\eV$), regardless of whether the adjacent carbon
atoms are already hydrogenated (see Figure
\ref{fig:CINEB-adsorp}). This finding suggests that, at
least for the pristine graphene surface considered here,
using coverage independent adsorption rate constants is
reasonable.

A second, even more important justification for using
coverage independent parameters lies in the kinetic behavior
at low temperatures. KMC simulations show that, when two H
atoms are present on the graphene surface, their association
(\chem{H_2} formation) is extremely fast once they come
within reaction distance, because the association barrier is
low (or effectively barrierless on graphene with NQEs) at
temperatures below 50~K. Consequently, the surface coverage
of H atoms remains very low under the simulated conditions
below 50~K. H atoms are consumed via \chem{H_2} formation
almost as soon as they encounter each other, so no
accumulation of H atoms occurs that would lead to high local
coverage. Therefore, even if coverage dependent barriers
existed, the system rarely explores configurations with
multiple H atoms in close proximity. The coverage
independent assumption is thus dynamically consistent for
the low temperature regime we study. Only when NQEs are not
considered, the two-H association, diffusion and desorption
are all extremely slow, the local coverage is very high.

\begin{figure*}[ht!]
\centering  
\includegraphics[width=0.7\linewidth]{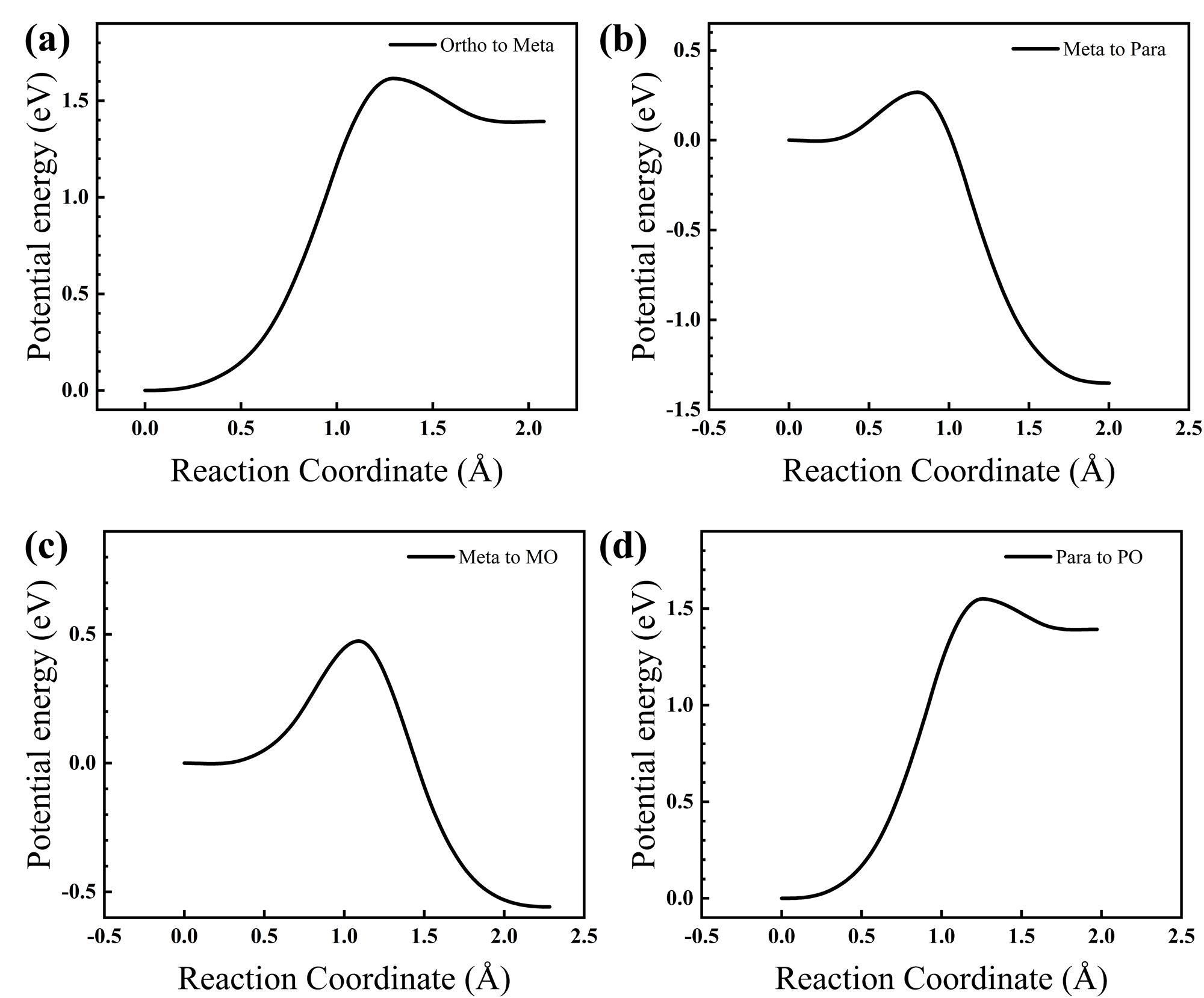}
\caption{Diffusion barriers of H atoms in dimer structures
  on the graphene surface for the transitions: (a) ortho to
  meta dimer, (b) meta to para dimer, (c) meta to MO dimer,
  and (d) para to PO dimer. These results are obtained using
  our DFT settings. 
  \label{fig:CINEB-hop}}
\end{figure*}

Previous literature reported that the hopping (diffusion)
barrier of H on graphene can be influenced by local coverage
and the presence of nearby H atoms \citep{Tong2024JPCC}. To
validate our computational setup, we perform calculations of
the hopping barrier under different local configurations and
compared them with published values. Our results agree very
well with the literature, confirming the reliability of our
DFT parameters. We consider only H hopping between two
adjacent sites, and present our results in Figure
\ref{fig:CINEB-hop}. Importantly, even when the hopping
barrier is affected by the local environment, its value
remains considerably higher than the adsorption and
association barriers in the interested temperature range. At
low temperatures, the hopping process is essentially frozen
on the timescale of \chem{H_2} formation, meaning that H
atoms do not migrate across the surface to form
high-coverage clusters. This further supports the validity
of using coverage independent parameters.

For the enstatite surface, we model \chem{H_2} formation
with a quasi-one-dimensional chain structure (discussed in
Appendix~\ref{sec:model}), and the distance between
different adsorption sites is large ($\sim
2.91~\ang$). Besides, we note that on the \chem{MgSiO_3}
surface, our calculations show that H adsorption is
barrierless even for the first H atom. Under these
conditions, coverage effects are expected to be less
critical than on carbonaceous surfaces. So, We do not
perform coverage dependent testing.

\subsection{Vacancy defect effects on the graphene surface}\label{sec:defect}

We first construct a $4\times 4$ supercell of graphene with
one missing C atom (a monovacancy), leaving three dangling
bonds. We calculate the potential energy profile of H
adsorption on a nearby carbon atom and find it to be
barrierless, releasing a large amount of energy
($\sim 4.2~\eV$) as the H atom saturates the dangling
bond. This indicates that H atoms should be used to saturate
carbon atoms adjacent to a vacancy. Furthermore, we
calculate the potential energy profile of an extra H
adsorption on a H saturated carbon atom near a vacancy,
which again is barrierless and releases $\sim 2.7~\eV$ of
energy. This likely occurs because, after vacancy formation,
adjacent carbon atoms can achieve $\mathit{sp^3}$
hybridization upon saturation by extra H atoms, which is
more stable than the original $\mathit{sp^2}$
hybridization. Our results are consistent with those of
\cite{2017Gran}. Moreover, previous studies have also shown
that \chem{H_2} molecules can spontaneously dissociate on
graphene surface with vacancy defects, providing H atoms to
saturate dangling bonds \citep{2017Gran}.

\begin{figure}
\centering  
\includegraphics[width=0.9\linewidth]{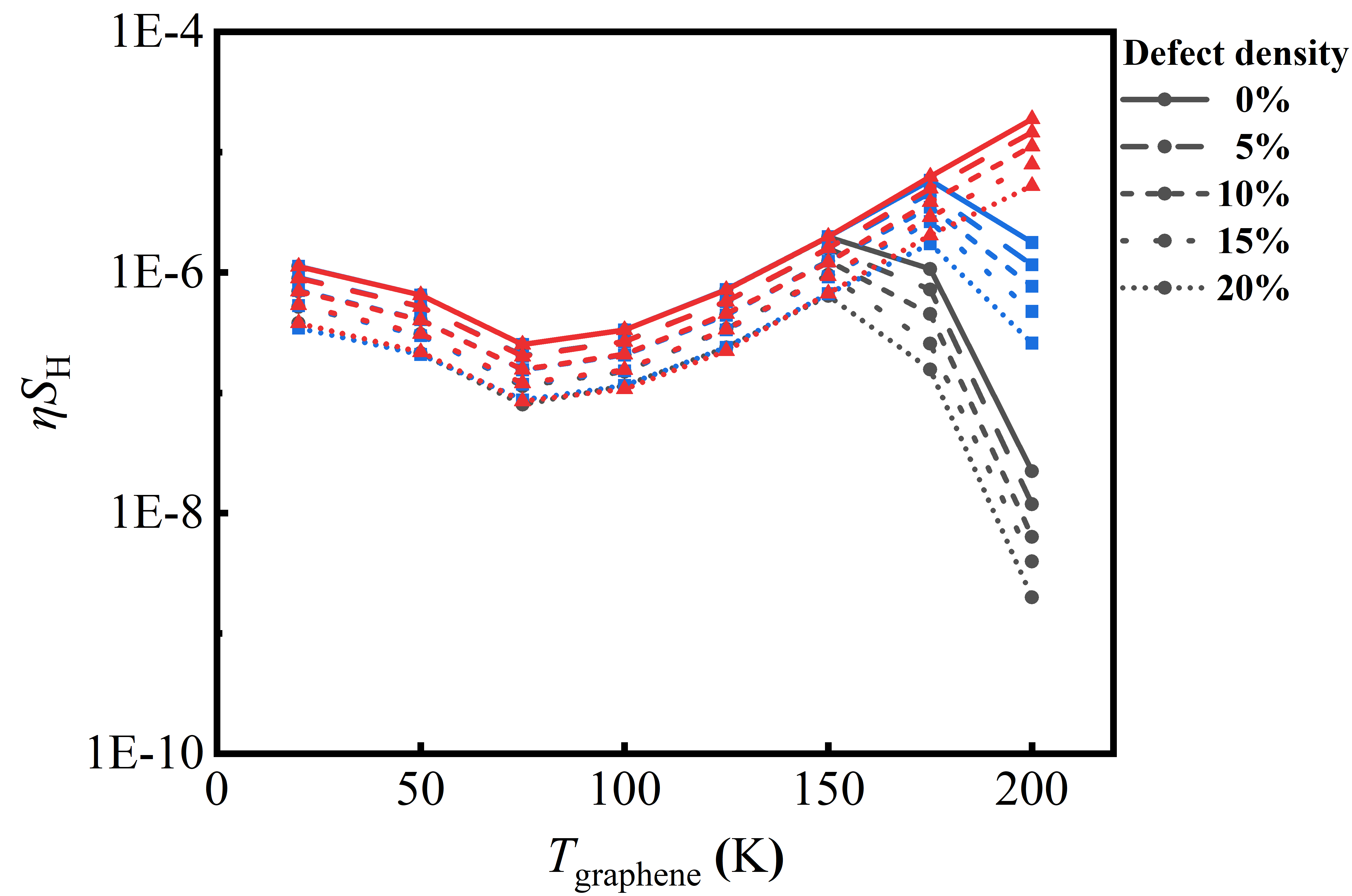}
\caption{\chem{H_2} formation efficiency coefficients versus
  $T_{\mathrm{dust}}$ on the graphene surface under the
  thermalization assumption for H adsorption at different
  vacancy densities and hydrogen densities of  
  $10^2~\cm^{-3}$ (black), $10^4~\cm^{-3}$ 
  (blue), and $10^{6}~\cm^{-3}$ (red). The vacancy density
  is defined as the ratio of the number of saturated C atoms
  to the total number of C atoms. 
  \label{fig:defect}}
\end{figure}

Once the dangling bonds are saturated, the saturated H atoms
face high energy barriers for both migration to neighboring
carbon atoms and recombination to form \chem{H_2}
molecules. These DFT results indicate that the carbon atoms
surrounding the vacancy are passivated by hydrogen atoms;
therefore, we assume that these sites no longer participate
in \chem{H_2} formation during the KMC simulations. Under
this assumption, we carried out new KMC simulations with
varying vacancy densities, where the vacancy density is
defined as the ratio of the number of saturated C atoms to
the total number of C atoms, and the results are shown in
Figure \ref{fig:defect}. As the vacancy density increases,
the \chem{H_2} formation efficiency gradually decreases
because more sites become passivated due to the strong
stability (or inertness equivalently) of $\mathit{sp^3}$ C
atoms around C vacancies.

We do not introduce defects on the \chem{MgSiO_3} surface
due to its structural complexity and the computational cost
of modeling disordered silicate lattices. A
  complete treatment of local environment effects, including
  coverage-dependent barriers, surface defects, and
  amorphous structures, is essential for a truly realistic
  simulation of interstellar grain surface chemistry. The
  number of possible local configurations, nonetheless,
  grows combinatorially, making brute-force parametrization
  for each distinct environment computationally
  prohibitive. Our multiscale framework could still be
  suited to address this challenge with practical
  approximations. For example, representative local motifs
  (steps, vacancies, adatoms) \citep[e.g.][]{2005Cuppen,
    2006Cuppen} can be treated with PIMC to obtain rigorous,
  temperature-dependent rate constants, while larger
  structural ensembles can be handled using potential energy
  barriers obtained via CI-NEB accelerated by MLFFs
  \citep{2026Guobing}, combined with harmonic TST and
  semiclassical tunneling corrections. Looking further
  ahead, we envisage that neural-network models could be
  trained to map local atomic environments directly to
  free-energy barriers or rate constants, bypassing the need
  for explicit PIMC or even CI-NEB calculations at every
  site. Such a surrogate would enable rapid evaluation of
  rate constants across the vast configurational space of
  amorphous grains, supplying input for large-scale KMC
  simulations of disordered surfaces with full coverage
  dependence without sacrificing the first-principles
  quality of the training set.

\subsection{The average hydrogen occupancy in KMC
  simulations}
We also compute the average hydrogen occupancy at surfaces
of these two materials under different conditions in our KMC
simulations. The hydrogen occupancy is determined by the
dynamic competition among adsorption, desorption and
association elementary steps. When the adsorption rate is
several orders of magnitude faster than those of the other
two steps, the occupancy would approach to 1. Occupancy
results are shown in Table \ref{tab:ocp_thermalization} (the
thermalization limit) and Table \ref{tab:ocp_adiabatic} (the
adiabatic limit). 

\begin{deluxetable*}{ccccccccc}[h]
  \tabletypesize{\scriptsize} \tablewidth{0pt}
  \tablecaption{Hydrogen
    occupancy at the graphene and the \chem{MgSiO_3}
    surfaces under conditions of different temperatures and
    different hydrogen densities with the thermalization
    assumption. \label{tab:ocp_thermalization}} \tablehead{
    \colhead{} 
    &\colhead{} & \multicolumn{3}{c}{Graphene}& \colhead{} &
    \multicolumn{3}{c}{\chem{MgSiO_3}}
    \\
    \colhead{$T_{\rm dust}~(\K)$} &\colhead{} &
    \multicolumn{3}{c}{Density (cm$^{-3}$)}& \colhead{} &
    \multicolumn{3}{c}{Density (cm$^{-3}$)}
    \\
    \cline{3-5}\cline{7-9} \colhead{} &\colhead{} &
    \colhead{$10^2$} & \colhead{$10^4$} & \colhead{$10^6$} &
    \colhead{ } & \colhead{$10^2$} & \colhead{$10^4$} &
    \colhead{$10^6$} }
\startdata 
50 & { } & 0.994/{0.049} & 0.995/{0.049}  &  0.995/{0.049}  & & 0.999/{0.997} & 0.999/{0.999} & 0.999/{0.999} \\
100 & { } & 0.094/{0.049} & 0.094/{0.049} & 0.094/{0.049}	& & 0.999/{0.233}	& 0.999/{0.242} & 0.999/{0.611} \\
150 & { } & 0.066/{0.043} & 0.066/{0.049} & 0.083/{0.049} & & 0.233/{0.233} & 0.247/{0.233} & 0.673/{0.233} \\
200 & { } & 0.027/{0.003} & 0.057/{0.004} & 0.065/{0.042} & & 0.109/{0.001} & 0.230/{0.060} & 0.234/{0.224} \\
\enddata
\tablecomments{Classical results (without NQEs) are presented first, followed by quantum results (with NQEs).}
\end{deluxetable*}

\begin{deluxetable*}{ccccccccc}[h]
  \tabletypesize{\scriptsize} \tablewidth{0pt}
  \tablecaption{Hydrogen
    occupancy at the graphene and the \chem{MgSiO_3}
    surfaces under conditions of different temperatures and
    different hydrogen densities with the adiabatic
    assumption ($T_{\rm
      gas}=600~\K$). \label{tab:ocp_adiabatic}} \tablehead{
    \colhead{} 
    &\colhead{} & \multicolumn{3}{c}{Graphene}& \colhead{} &
    \multicolumn{3}{c}{\chem{MgSiO_3}}
    \\
    \colhead{$T_{\rm dust}~(\K)$} &\colhead{} &
    \multicolumn{3}{c}{Density (cm$^{-3}$)}& \colhead{} &
    \multicolumn{3}{c}{Density (cm$^{-3}$)}
    \\
    \cline{3-5}\cline{7-9} \colhead{} &\colhead{} &
    \colhead{$10^2$} & \colhead{$10^4$} & \colhead{$10^6$} &
    \colhead{ } & \colhead{$10^2$} & \colhead{$10^4$} &
    \colhead{$10^6$} }
\startdata 
50 & { } & 0.995/{0.049} & 0.995/{0.059}  &  0.995/{0.066}  &  & 0.999/{0.999} & 0.999/{0.999} & 0.999/{0.999} \\
100 & { } & 0.100/{0.049} & 0.360/{0.049} & 0.952/{0.049}	& & 0.999/{0.234}	& 0.999/{0.258} & 0.999/{0.756} \\
150 & { } & 0.092/{0.050} & 0.093/{0.049} & 0.093/{0.049} & & 0.233/{0.234} & 0.262/{0.232} & 0.780/{0.233} \\
200 & { } & 0.065/{0.026} & 0.066/{0.049} & 0.074/{0.062} & & 0.136/{0.002} & 0.231/{0.086} & 0.233/{0.228} \\
\enddata
\tablecomments{Classical results (without NQEs) are
  presented first, followed by quantum results (with NQEs).} 
\end{deluxetable*}

\subsection{Mechanism of hydrogen formation at dust grains with NQEs}

KMC simulations reveal that the \chem{H_2} formation
mechanism is strongly influenced by temperature, atomic
hydrogen density, and grain composition. On graphene below
50~K, the energy barriers for diffusion and desorption
remain significantly higher than those for adsorption and
two-H association, effectively suppressing hopping and
desorption. Consequently, only the latter two processes are
relevant while diffusion and desorption hardly happen. With
NQEs reducing the two-H association barrier to a negligible
value, adsorption emerges as the rate-limiting step for
hydrogen formation under the thermalization assumption. At
200~K, although two-H association remains the fastest step,
diffusion and desorption rates become competitive with (or
even exceed) the adsorption with NQEs under the
thermalization assumption. At high H densities (e.g.,
$10^6~\cm^{-3}$), adsorption continues to dominate over
desorption, and NQEs exhibit negligible influence on the
adsorption process (rate-limiting step), resulting in
negligible change in the \chem{H_2} formation rate with or
without NQEs. At low densities, however, adsorption becomes
slower than desorption and diffusion when NQEs are included,
which means a chemisorbed H atom is likely to desorb before
reacting, leading to a net decrease in the formation
rate. When switching from the thermalization assumption
($T_{\mathrm{gas}} = T_{\mathrm{dust}}$) to the adiabatic
assumption ($T_{\mathrm{gas}} \gg T_{\mathrm{dust}}$), the
\chem{H_2} formation rate increases greatly due to a much
higher sticking probability. On the \chem{MgSiO_3} surface,
which exhibits barrierless adsorption, the reaction behavior
differs. At low temperatures, two-H association is slower
than adsorption and becomes the rate-limiting step
(diffusion and desorption are much slower, thus being
excluded), leading to an accumulation of $\sim 1$ single
monolayer of \chem{H^*} coverage. Also, switching from the
thermalization assumption to the adiabatic assumption, the
\chem{H_2} formation rate remains unchanged. At 200~K,
association is the fastest and the \chem{H_2} formation rate
is determined by the competition between adsorption and
desorption. At low H densities, desorption also influences
the formation rate, showing a trend similar to that on
graphene at $10^2~\cm^{-3}$. At high H densities, adsorption
is faster than desorption, causing the formation rate to
approach its upper limit. Given that adsorption is
barrierless, switching to the adiabatic assumption has a
negligible effect on adsorption rate constant and thus on
the final \chem{H_2} formation rate.


\end{document}